\documentclass[nofootinbib]{revtex4}
\usepackage{fullpage}
\usepackage[dvips]{graphicx}
\input{epsf}
\usepackage{amsmath}
\usepackage{amssymb}
\usepackage[mathscr]{eucal}
\usepackage{bm}
\usepackage{theorem}

\newcommand{\beq}{\begin{equation}}
\newcommand{\eeq}{\end{equation}}
\newcommand{\beqa}{\begin{eqnarray}}
\newcommand{\eeqa}{\end{eqnarray}}
\newcommand{\beqan}{\begin{eqnarray*}}
\newcommand{\eeqan}{\end{eqnarray*}}

\newcommand{\de}[2]{ \frac{d #1}{d #2} }

\newcommand{\tr}[1]{{\rm tr} \left( #1 \right) }
\newcommand{\ket}[1]{| #1 \rangle}

\newcommand{\ketbra}[2]{| #1   \rangle \langle #2 |}

\newcommand{\op}{\oplus}

\newcommand{\llangle}{\langle\!\langle} 
\newcommand{\rrangle}{\rangle\!\rangle}

\newcommand{\bmrho}{\bm{\varrho}}
\newcommand{\bmlambda}{\bm{\lambda}}

\newcommand{\ad}{{\rm ad}}

\newcommand{\proof}{\noindent {\bf Proof. }}
\newcommand{\qed}{\hfill $\Box$ \vskip 2ex}

\newtheorem{theorem}{Theorem}

\newtheorem{proposition}{Proposition}
\newtheorem{lemma}{Lemma}
\newtheorem{corollary}{Corollary}

{\theorembodyfont{\upshape} 
\newtheorem{remark}{Remark}
\newtheorem{example}{Example}
}

\begin{document}

\title{Feedback stabilization of quantum ensembles: a global convergence analysis on complex flag manifolds}
\author{Claudio Altafini}
\affiliation{SISSA-ISAS  \\
International School for Advanced Studies \\
via Beirut 2-4, 34014 Trieste, Italy }
\email{altafini@sissa.it}


\begin{abstract}

In an N-level quantum mechanical system, the problem of unitary feedback stabilization of mixed density operators to periodic orbits admits a natural Lyapunov-based time-varying feedback design.
A global description of the domain of attraction of the closed-loop system can be provided based on a ``root-space''-like structure of the space of density operators.
This convex set foliates as a complex flag manifold where each leaf is identified with the coadjoint orbit of the eigenvalues of the density operator.
The converging conditions are time-independent but depend from the topology of the flag manifold: it is shown that the closed loop must have a number of equilibria at least equal to the Euler characteristic of the manifold, thus imposing obstructions of topological nature to global stabilizability.

\end{abstract}

\maketitle 


\section{Introduction}

The use of feedback in quantum mechanics is limited by the phenomenon of wavefunction collapse following a measurement.
In this work the problem is bypassed by considering density operators of quantum ensembles and completely noninvasive measurements.
This allows also to relax the requirement of commutativity of the measured observables and in fact we shall assume to have a complete knowledge of the density operator for all times.
Although physically this set up is realistic only for some applications (typically nuclear spin ensembles \cite{Cory1,Havel1}), it is of widespread use for the purposes of model-based quantum control (often under the name ``tracking control'' \cite{Brown1,Zhu1}), as it allows to generate control fields in spite of the high complexity of open loop control \cite{Boscain2,DAlessandro1,Schirmer5}.
Furthermore, while the formulation comes from quantum control, our motivations for this work are mostly mathematical, namely feedback design and convergence analysis for a class of bilinear control systems living on a particular family of compact manifolds and evolving isospectrally.
As the system has a drift term which cannot be canceled without incurring in singularities of the control law, the most natural problem formulation is to seek for a stabilizer to the periodic orbit drawn by the drift.
Rather than studying this problem like an orbital stabilization problem \cite{Bacciotti1}, we reformulate and solve it as a state tracking problem, thus avoiding the obstruction to semiglobal convergence of a periodic orbit, see \cite{Wilson1}, Corollary 1.6 (where it is called stability in the large).
In fact, with our feedback design the state will converge to the orbit, but the entire orbit is not an invariant set, only a point moving along it is invariant.
As a matter of fact, by passing to a suitable rotating frame, our time-dependent trajectory tracking problem can be reformulated completely in terms of time-varying feedback law for the fixed point of a nonautonomous system.
The Lyapunov design is essentially of the Jurdjevic-Quinn type \cite{Jurdjevic5}, for which the usual LaSalle invariance principle is applicable in spite of the time-dependence of the closed loop, and does not differ much from what has already been proposed in the literature for wavefunctions \cite{Ferrante1,Vettori1,Grivopoulos1,Mirrahimi1}.

What is nontrival is to ascertain the convergence of certain initial conditions and to provide a global description of the region of attraction.
In fact, the sufficient condition used in \cite{Jurdjevic5} to prove asymptotic convergence and based on the so-called $\ad$-condition or Jurdjevic-Quinn condition \cite{Bacciotti2}, is never verified globally for $ N>2$.
This is due to the presence of an abelian subalgebra (Cartan subalgebra) that can never be fully spanned by $ \ad$-commutators alone.
It will be shown, however, that the undesired critical points are not only unstable but also repulsive, meaning that the Jurdjevic-Quinn condition or, equivalently, the controllability of the linearization  along the desired orbit \cite{Mirrahimi1}, guarantees convergence for all initial conditions outside the set of equilibrium points.
To attain a complete and time-independent description of the critical set and thus of the domain of attraction, a thorough geometric and topologic characterization of the state manifold of mixed density operators is required.
A unitary evolution like that appearing in a Liouville equation is isospectral, as the eigenvalues of the density operator form a complete set of invariants.
Unlike for a wavefunction, the state space has dimension and structure which depend on the multiplicities of such eigenvalues.
Since these form a flag in dimension $N$, all complex flag manifolds obtained as homogeneous spaces of $ U(N)$ (or $ SU(N)$) by the Cartesian products of subgroups of dimensions given by the multiplicities are admissible state spaces \cite{Bengtsson1,Picken1,Zyczkowski2}.
Since the set of density operators up to the imaginary unit ``overlaps'' with the Lie algebra $ \mathfrak{u}(N) $ (or $ \mathfrak{su}(N)$, excluding the constant trace), these complex flag manifolds can also be intended as the orbits of the (co)adjoint action of $ U(N)$ (or $SU(N)$) on its Lie algebra.
This is a well-studied action and the structure of its orbits is well-known \cite{Marsden2,Frankel1}: for example a fundamental topological invariant like the Euler characteristic acquires the meaning of number of nontrivial possible permutations of the eigenvalues of the density operator, see also \cite{Bengtsson1,Chaturvedi1,Ercolessi1,Zyczkowski2}.
For the purposes of stabilizability, this is an important feature, because it will be shown that each complex flag manifold has a number of ``antipodal'' points equal to the Euler characteristic, and that these points must be equilibria of the closed loop system.
In order to give a complete description of the region of attraction, we use the resemblance between the set of density operators and the Lie algebra $ \mathfrak{su}(N)$, and a few tools deriving from the root space decomposition of a compact Lie algebra, namely its orthogonal decomposition into Cartan subalgebra plus root spaces and the invariance properties of the root spaces under certain commutators (like the ad-commutators) \cite{Cla-contr-root1}.
This ``graph-like'' approach yields simple, time-independent characterizations of all converging initial conditions for a given reference orbit and Hamiltonian.
Also the Kalman controllability of the linearization admits an intrinsic formulation in these terms.
The formalism used gives insight into the problem of choosing reference orbits having a large domain of attraction.


It is known \cite{Bhat1,Koditschek3}, that compact manifolds do not admit a global asymptotically stable equilibrium because they are not contractible.
This is a topological property and corresponds to a set being homotopy equivalent to a point \cite{Guillemin1}.
The region of convergence of an asymptotically stable attractor must be in such a homotopy class \cite{Bhat1,Wilson1}.
For our complex flag manifolds, it will be shown that the antipodal points represent topological obstructions to global stabilizability.

In order to simplify the treatment, an equivalent real representation of density operators is used throughout, given by the so-called coherence vector and corresponding to the vector of expectation values with respect to a complete orthonormal set of Hermitian matrices \cite{Alicki1,Bengtsson1,Schirmer7}.
It provides a linear representation of the adjoint action occurring in a Liouville equation \cite{Cla-spin-tens1}, and it allows to formulate the control system in terms of standard bilinear systems on smooth manifolds which are real representations of the complex flag manifolds.

\section{Driven Liouville-von Neumann equation}

With a given Hamiltonian $ H = H_A + u H_B $, $ -i H_A , \, -i H_B \in \mathfrak{su}(N) $, $ u \in C^\infty (\mathbb{R} )$ a control field, one can form a Schr\"{o}dinger equation for the wavefunction $ \ket{\psi} $ (in atomic units, $\hbar=1$)
\begin{equation}
  {\ket{\dot \psi }} = -i \left(  H_A + u \,  H_B \right)
\ket{\psi }  , \qquad \ket{\psi(t) } \in \mathbb{S}^{2N-1} ,
\label{eq:schrod1}
\end{equation}
or a Liouville-von Neumann equation for the density operator $ \rho $
\beq
\dot \rho = - i [ H_A + u \, H_B , \, \rho ] , \qquad \rho=\rho^\dagger \geqslant 0 , \quad {\rm tr}(\rho)=1, \quad {\rm tr}(\rho^2) \leqslant 1.
\label{eq:Liouville1}
\eeq
Eq.~\eqref{eq:Liouville1} holds for a quantum ensemble, hence it is more general than \eqref{eq:schrod1}; the two being equivalent only when $ {\rm tr}(\rho^2) =1 $ (i.e., $ \rho $ is a rank-one operator: $\rho = \ketbra{\psi}{\psi} $).

\subsection{Gell-Mann basis and adjoint representation}

The left hand side of \eqref{eq:Liouville1} contains a conjugation action on state matrices.
In order to deal with it, we can use one of the features of the adjoint representation, namely its providing a linear representation of one-parameter groups of automorphisms of $ \mathfrak{su}(N) $ (see Appendix~\ref{sec:adj-rep}), and reformulate \eqref{eq:Liouville1} as a standard bilinear control systems on a suitable manifold.
To do that, we use the coherence vector representation of $ \rho $, whose key property is that both $ H $ and $ \rho $ are expressed in terms of the same complete orthonormal set of Hermitian matrices.
Let $ \lambda_0 = \frac{1}{\sqrt{N}} \openone_N $ and call $ {\bm \lambda} $ the $ n$-dimensional vector of $ N \times N $ Gell-Mann matrices \cite{Georgi1}, $ n= N^2 -1 $.
Since $ \mathfrak{su}(N) $ contains traceless skew-hermitian matrices, $ {\rm span } \{ -i {\bm \lambda}  \} = \mathfrak{su}(N) $.
Denote with $ \mathfrak{h} $ the Cartan subalgebra of $ \mathfrak{su}(N)$, i.e., the maximally abelian subalgebra in $ \mathfrak{su}(N)$, $ {\rm dim} (\mathfrak{h}) = N-1 $.  
In the Gell-Mann basis $ {\bm \lambda} $, $ \mathfrak{h} $ corresponds to the $N-1 $ diagonal matrices.
Denote $ \mathfrak{k} $ the vector space such that $ \mathfrak{su}(N) = \mathfrak{h} \op \mathfrak{k} $, with $ \mathfrak{h} \perp  \mathfrak{k} $ in a standard biinvariant $\mathfrak{su}(N) $ metric (e.g. the Killing metric).
While $ \mathfrak{h} $ is an abelian subalgebra, $  \mathfrak{k} $ is only a vector space.
In correspondence of this direct sum, we have the decomposition of $ {\bm \lambda} $ into $ \bmlambda_ \mathfrak{h} $ and $  \bmlambda_ \mathfrak{k} $ so that $\mathfrak{h} = {\rm span } \{ -i  \bmlambda_ \mathfrak{h} \} $ and $\mathfrak{k} = {\rm span } \{ -i  \bmlambda_ \mathfrak{k} \} $. 
If $ E_{j\ell} $ is the elementary $N\times N $ matrix having 1 in the $ (j\ell)$ slot and 0 elsewhere, then the matrices $ {\bm \lambda }$ are given by
\beq
\left\{ \lambda_{\mathfrak{h}, j}, \quad 1\leqslant j\leqslant  N-1  \right\}  =  \left\{ (E_{11} + \ldots + E_{jj}- j E_{j+1,j+1})/\sqrt{j(j+1)}, \quad 1\leqslant j\leqslant N-1 \right\} \label{eq:basis-elem1} 
\eeq
for the diagonal part, and
\beqa
\left\{ \lambda_{\mathfrak{k}, \Re, j \ell } , \quad 1 \leqslant j<\ell \leqslant N \right\} & = & \left\{ ( E_{j\ell}+E_{\ell j})/\sqrt{2}, \quad 1 \leqslant j<\ell \leqslant N \right\} \label{eq:basis-elem3} \\
\left\{ \lambda_{\mathfrak{k}, \Im, j \ell } , \quad 1\leqslant j<\ell\leqslant N \right\} & = & \left\{ i ( - E_{j\ell}+E_{\ell j})/\sqrt{2}, \quad 1\leqslant j<\ell\leqslant N \right\} \label{eq:basis-elem4}
\eeqa
for the off-diagonal part.
Calling $ \mathfrak{k}_{j\ell} = {\rm span} \left\{ -i \lambda_{\mathfrak{k}, \Re, j \ell }, -i  \lambda_{\mathfrak{k}, \Im, j \ell }  \right\}  $, then we have the further splitting of $ \mathfrak{k}$ into ``root spaces'' 
\beq
\mathfrak{k} = \bigoplus_{1\leqslant j<\ell\leqslant N } \mathfrak{k}_{j\ell} 
\label{eq:root-spaces1}
\eeq 
with the following commutation relations (see for instance \cite{Cla-contr-root1} for the details):
\beq
[ \mathfrak{h}, \, \mathfrak{k}_{j\ell} ]  =  \mathfrak{k}_{j\ell}, \label{eq:Cartan-sub-brack1}  
\eeq
\beq
[  \mathfrak{k}_{j\ell}, \,  \mathfrak{k}_{pq} ] =  \begin{cases}
\emptyset & \text{ if $ \ell\neq p $ and $ j \neq q $} \\
 \mathfrak{k}_{jq} & \text{ if $  \ell = p $} \\
 \mathfrak{k}_{p\ell} & \text{ if $ j = q $} \\
\subseteq  \mathfrak{h} & \text{ if $ j = p $ and $ \ell = q $} .
\end{cases}
\label{eq:Cartan-sub-brack2}
\eeq

Assume $ H_A $ is diagonal
\[
H_A = \begin{bmatrix} 
{\cal E}_1\\ &   \ddots & \\  & & {\cal E}_N 
\end{bmatrix}, \qquad {\cal E}_1 + \ldots + {\cal E}_N =0,
\]
and nondegenerate, i.e., $ {\cal E}_j \neq {\cal E}_\ell $, $ j\neq \ell $, where the $ {\cal E}_j $ are supposed ordered: ${\cal E}_1 < {\cal E}_2 < \ldots < {\cal E}_N $.
The $ {\cal E}_j \in \mathbb{R} $ are the energy levels of the (unforced) system \eqref{eq:schrod1}, i.e., the eigenvalues of the stationary Schr\"{o}dinger equation $ H_A \ket{\psi_j} = {\cal E}_j \ket{\psi_j}  $ of eigenstates $ \ket{\psi_j} =  {\rm e}_j  $ ($ {\rm e}_j $ is the elementary basis vector), $ j=1, \ldots, N$.
Assume also that the transition frequencies are nondegenerate, i.e., that the levels are not equispaced $ {\cal E}_j - {\cal E}_\ell \neq {\cal E}_p -{\cal E}_q  $, $ ( j\ell ) \neq (pq) $ $ j\neq \ell $, $ p\neq q $. 
Further standard assumptions are that $ H_B $ is off-diagonal and $ {\rm Graph}( H_B ) $ connected.
A stronger assumption we shall need is that $ H_B $ enables all transitions among adjacent energy levels: $ \tr{ H_B \mathfrak{k}_{j,j+1} } \neq 0$ $ \forall \; j=1, \ldots, N-1 $.
Beside connectivity of $ {\rm Graph}( H_B ) $, it guarantees that all ``fundamental root spaces'' (see \cite{Cla-contr-root1}) are excited by the dynamics.
Since $ H_A $ is diagonal and traceless, $ -i H_A \in \mathfrak{su}(N)$.
A nondegenerate element of the Cartan subalgebra (like $ H_A $ above) is called regular. 
It is called strongly regular if in addition it has all nondegenerate transitions.

In terms of $ {\bm \lambda} $, $ H_A = {\bm h}_A \cdot {\bm \lambda} $ and $ H_B = {\bm h}_B \cdot {\bm \lambda} $, with $ {\bm h}_A ,\,  {\bm h}_B \in \mathbb{R}^n$.
Likewise $ \rho = \varrho_0 \lambda_0 + \bmrho \cdot {\bm \lambda} $, where $ \varrho_0 = \frac{1}{\sqrt{N}} $ and the coherence vector $ \bmrho $ is composed of expectation values along the basis elements \eqref{eq:basis-elem1}-\eqref{eq:basis-elem4}: $ \varrho_{\mathfrak{h},j} = {\rm tr} ( \rho \lambda_{\mathfrak{h},j} )\in \mathbb{R}$, $ j=1, \ldots, N-1 $, $ \varrho_{\mathfrak{k},\Re,j\ell } = {\rm tr} ( \rho \lambda_{\mathfrak{k},\Re,j\ell} )\in \mathbb{R}$, $ \varrho_{\mathfrak{k},\Im,j\ell } = {\rm tr} ( \rho \lambda_{\mathfrak{k},\Im,j\ell} )\in \mathbb{R}$, $ 1\leqslant j<\ell\leqslant N$.
Any density $ \rho $ can be split as $ \rho =  \varrho_0 \lambda_0 +\rho_ \mathfrak{h} + \rho_\mathfrak{k} $, or in correspondence of \eqref{eq:basis-elem1}-\eqref{eq:basis-elem4}, $  \rho = \varrho_0 \lambda_0 + \bmrho_ \mathfrak{h} \cdot \bmlambda_ \mathfrak{h} +  \bmrho_ \mathfrak{k}\cdot \bmlambda_ \mathfrak{k}= \varrho_0 \lambda_0 + \sum_{1\leqslant j < N} \varrho_{\mathfrak{h}, j} \lambda_{\mathfrak{h}, j} +  \sum_{1\leqslant j < \ell \leqslant N} \left( \varrho_{\mathfrak{k},\Re, j \ell} \lambda_{\mathfrak{k},\Re, j\ell } + \varrho_{\mathfrak{k},\Im, j \ell} \lambda_{\mathfrak{k},\Im, j\ell } \right) $.
The nonzero components of the coherence vector $ \bmrho $ uniquely identify a subset of the $  \mathfrak{k}_{j\ell} $.
Denote $ \mathfrak{f}_\mathfrak{k} (\rho) $ the ``support'' of $ \rho $ in $ \mathfrak{k} $, i.e., the set of root spaces ``touched'' by $ \rho $: $ \mathfrak{f}_\mathfrak{k} (\rho) = \rho  \cap  \mathfrak{k} $.
Also let $ {\cal F}_\mathfrak{k} (\rho)  =\{ (j\ell) \text{ s.t. }  \tr{\rho  \mathfrak{k}_{j\ell}} \neq 0  , \; 1\leqslant j<\ell\leqslant N  \} $ be the corresponding set of index pairs. 
When $ (j\ell ) \in {\cal F}_\mathfrak{k} ( \rho) $, then $( \varrho_{\mathfrak{k},\Re, j \ell}, \varrho_{\mathfrak{k},\Im, j\ell }) \neq ( 0, 0) $.
Likewise $ \mathfrak{f}_\mathfrak{h} (\rho) = \rho  \cap  \mathfrak{h} $ and ${\cal F}_\mathfrak{h} (\rho)= \{ (j) \text{ s.t. }  \varrho_{\mathfrak{h}, j } \neq 0  , \; 1\leqslant j< N  \} $.
In the following we shall use both symbols $ \rho $ and $ \bmrho $ for densities and we shall refer to ``diagonal'' and ``off-diagonal'' $ \bmrho $ with an obvious abuse of notation.
The Hilbert-Schmidt norm on density operators induces for $ \bmrho $ the standard Euclidean norm $ \| \, \cdot \, \| $: $ \tr{\rho^2} =\varrho_0^2+ \llangle \bmrho,\bmrho \rrangle = \varrho_0^2+\| \bmrho \| ^2 $, where we indicate with $ \llangle \, \cdot \, , \, \cdot \, \rrangle $ the $ \mathbb{R}^n $-Euclidean inner product.
Due to the trace-class constraint, the notion of distance between the densities $ \rho_1 $ and $ \rho_2$ having the same purity is $ d(\rho_1 , \rho_2 ) = \tr{\rho_1^2} - \tr{\rho_1 \rho_2}$, see e.g. \cite{Zyczkowski2}, or in terms of $ \bmrho$: 
\beq
d(\bmrho_1 , \bmrho_2 ) = \varrho_0^2+\| \bmrho_1 \| ^2 - \varrho_0^2 - \llangle \bmrho_1, \bmrho_2 \rrangle 
= \| \bmrho_1 \| ^2 - \bmrho_1^T  \bmrho_2 \; \in [ 0, \, \tr{\rho_1^2} ] .
\label{eq:dist-cohvec}
\eeq
Thanks to the use of the same basis for $ \rho $ and the Hamiltonian, up to the imaginary unit the trajectories of \eqref{eq:Liouville1} can be identified with the adjoint orbits of $SU(N)$ on its Lie algebra, see Section~\ref{sec:topol-descr} for a thorough description.
Following Appendix~\ref{sec:adj-rep}, we can replace the matrix ODE \eqref{eq:Liouville1} with the linear vector ODE 
\beq
\dot \bmrho = \left( {\bm A} + u \, {\bm B} \right) \bmrho  ,
\label{eq:Liouv-vect-coh1}
\eeq
where $ {\bm A} = -i \ad_{H_A} = -i {\bm h}_A \cdot \ad_{\bm \lambda}  $, $ {\bm B} = -i \ad_{H_B} = -i {\bm h}_B \cdot \ad_{\bm \lambda} $ and $ {\bm A} ,{\bm B}\in \ad_{\mathfrak{su}(N)} \subset \mathfrak{so}(n)$.
The Lie algebra $ \ad_{\mathfrak{su}(N)} $ is the adjoint representation of $ \mathfrak{su}(N) $, hence $ {\rm dim} (\ad_{\mathfrak{su}(N)}) = {\rm dim} (\mathfrak{su}(N)) = n $.
Therefore, for $ N> 2 $, $ \ad_{\mathfrak{su}(N)} \subsetneq \mathfrak{so}(n)$.

The isomorphism $ \mathfrak{su}(N) \simeq \ad_{\mathfrak{su}(N)}$ induces an orthogonal splitting also in the adjoint representation: $ \ad_{\mathfrak{su}(N)}=\ad_{ \mathfrak{h}} \op \ad_{\mathfrak{k}}$.
Obviously relations similar to \eqref{eq:Cartan-sub-brack1}-\eqref{eq:Cartan-sub-brack2} still hold and $ {\bm A} = -i {\bm h}_{A, \mathfrak{h}} \cdot \ad_{\bmlambda _ \mathfrak{h}} $, $ {\bm B} = -i {\bm h}_{B, \mathfrak{k}} \cdot \ad_{\bmlambda _ \mathfrak{k}} $.
For $ C \in \mathfrak{su}(N) $, in components $  C =  \sum_{1\leqslant j < N} c_{\mathfrak{h}, j} \lambda_{\mathfrak{h}, j} +  \sum_{1\leqslant j < \ell \leqslant N} \left( c_{\mathfrak{k},\Re, j \ell} \lambda_{\mathfrak{k},\Re, j\ell } + c_{\mathfrak{k},\Im, j \ell} \lambda_{\mathfrak{k},\Im, j\ell } \right) $, we also shall indicate with $ \mathfrak{f}_\mathfrak{h} (C) $, $ \mathfrak{f}_\mathfrak{k} (C) $ the support of $ C$ in, respectively, $ \mathfrak{h} $, $ \mathfrak{k}$, of indices ${\cal F}_\mathfrak{h} (C) $, ${\cal F}_\mathfrak{k} (C)$.

For later use, we need to compute some of the commutators of \eqref{eq:Cartan-sub-brack1}-\eqref{eq:Cartan-sub-brack2} more in detail.
\beq
[ \lambda_{\mathfrak{k} , \Re, j\ell } , \,  \lambda_{\mathfrak{k} , \Im, j\ell } ] = i ( E_{jj} - E_{\ell \ell } ) = \begin{cases}
- \sqrt{\frac{j-1}{j}} \lambda_{\mathfrak{h} , j-1 } + \sum_{p=j}^\ell \frac{1}{\sqrt{p(p+1)}} \lambda_{\mathfrak{h} ,p } +\sqrt{\frac{\ell}{\ell-1}} \lambda_{\mathfrak{h} , \ell -1 } & \text{ if $ j>1 $ and $ \ell > 2 $} \\
\sum_{p=j}^\ell \frac{1}{\sqrt{p(p+1)}} \lambda_{\mathfrak{h} ,p } +\sqrt{\frac{\ell}{\ell-1}} \lambda_{\mathfrak{h} , \ell -1 } & \text{ if $ j=1 $ and $ \ell > 2 $} \\
\sqrt{\frac{\ell}{\ell-1}} \lambda_{\mathfrak{h} , \ell -1 } & \text{ if $ \ell =2  $} 
\end{cases}
\label{eq:comm-h1}
\eeq
For $ (j\ell ) \neq ( pq) $:
\beq
\begin{split}
[ \lambda_{\mathfrak{k} , \Re, j\ell } , \,  \lambda_{\mathfrak{k} , \Re, pq } ] & = 
\frac{i}{\sqrt{2}} \left( 
 \delta_{\ell p}  \lambda_{\mathfrak{k} , \Im, jq } 
+\delta_{jp}  \lambda_{\mathfrak{k} , \Im,\ell q } 
+\delta_{jq}  \lambda_{\mathfrak{k} , \Im,\ell p } +
\delta_{lq}  \lambda_{\mathfrak{k} , \Im,jp } \right)
\\
[ \lambda_{\mathfrak{k} , \Re, j\ell } , \,  \lambda_{\mathfrak{k} , \Im, pq } ] & = 
\frac{i}{\sqrt{2}} \left( 
- \delta_{\ell p}  \lambda_{\mathfrak{k} , \Re, jq } 
-\delta_{jp}  \lambda_{\mathfrak{k} , \Re,\ell q } 
+\delta_{jq}  \lambda_{\mathfrak{k} , \Re,\ell p } +
\delta_{lq}  \lambda_{\mathfrak{k} , \Re,jp } \right)
\\
[ \lambda_{\mathfrak{k} , \Im, j\ell } , \,  \lambda_{\mathfrak{k} , \Im, pq } ] & = 
\frac{i}{\sqrt{2}} \left( 
- \delta_{\ell p}  \lambda_{\mathfrak{k} , \Im, jq } 
+\delta_{jp}  \lambda_{\mathfrak{k} , \Im,\ell q } 
-\delta_{jq}  \lambda_{\mathfrak{k} , \Im,\ell p } +
\delta_{lq}  \lambda_{\mathfrak{k} , \Im,jp } \right)
\end{split}
\label{eq:comm-kk}
\eeq

\subsection{Unforced equation}

For pure states in an orthonormal basis, the eigenvectors $ \ket{\psi_j}={\rm e}_j $ of the stationary Shr{\"o}dinger equation are mapped into the diagonal density operator $ \ketbra{\psi_j}{\psi_j}= E_{jj} $. More generally, for quantum ensembles, after a suitable diagonalization, $ \tilde{\rho} = {\rm diag} ( w_1 , \ldots , w_N)$, $ 0\leqslant w_j \leqslant 1 $, $ \sum_{j=1}^N w_j =1 $. 
The eigenvalues $ w_j $ represent the populations of the various energy levels and provide a complete set of invariants for \eqref{eq:Liouville1}, call it $ {\cal J} = \{ w_1, \ldots , w_N \} $, since \eqref{eq:Liouville1} is isospectral.

\begin{proposition}
\label{cor:diag-comm1}
Consider the system \eqref{eq:Liouv-vect-coh1} with $H_A $ strongly regular.
The state $ \bmrho $ is an equilibrium point of \eqref{eq:Liouv-vect-coh1} for $ u=0$ if and only if $ \rho = \varrho_0 \lambda_0 + \bmrho_\mathfrak{h} \bmlambda_\mathfrak{h} $.
Furthermore, if $ \bmrho_\mathfrak{k} \neq 0 $, then for $ u=0$
\begin{enumerate}
\item \label{item:drift1} $ \mathfrak{f}_\mathfrak{k} (\rho (0) ) = \mathfrak{f}_\mathfrak{k} (\rho (t) )$;
\item \label{item:drift2} $ \varrho_{\mathfrak{k}, \Re, j\ell } ^2 + \varrho_{\mathfrak{k}, \Im, j\ell } ^2 = {\rm const} $;
\item \label{item:drift3} for $ \delta t $ small, $ \varrho_{\mathfrak{k}, \Re, j\ell } (t) \neq  \varrho_{\mathfrak{k}, \Re, j\ell } (t + \delta t ) $ and $ \varrho_{\mathfrak{k}, \Im, j\ell } (t) \neq  \varrho_{\mathfrak{k}, \Im, j\ell } (t + \delta t ) $ $ \forall \; ( j \ell ) \in {\cal F}_\mathfrak{k} ( \rho ) $.
\end{enumerate}
\end{proposition}

\proof
When $ u=0$, for a given $ \rho = \varrho_0 \lambda_0 + \rho_\mathfrak{h} + \rho_\mathfrak{k}$, 
\beqa 
-i [ H_A , \, \rho_\mathfrak{h} ] & = & 0 \label{eq:adH_Arho0} \\
-i [ H_A , \, \rho ] & = & -i [ H_A , \, \rho_\mathfrak{k}  ] = -i [ H_A , \, \sum_{(j\ell) \in {\cal F}_\mathfrak{k} ( \rho ) }  \varrho_{\mathfrak{k},\Re,j \ell } \lambda_{\mathfrak{k},\Re,j \ell } +  \varrho_{\mathfrak{k},\Im,j \ell } \lambda_{\mathfrak{k},\Im,j \ell }]\nonumber  \\
& = & \sum_{(j\ell) \in {\cal F}_\mathfrak{k} ( \rho ) } 
\left( {\cal E}_j - {\cal E}_\ell \right) 
\left(   \varrho_{\mathfrak{k},\Re,j \ell }\lambda_{\mathfrak{k},\Im,j \ell } - \varrho_{\mathfrak{k},\Im,j \ell } \lambda_{\mathfrak{k},\Re,j \ell } \right).
\label{eq:adH_Arho1}
\eeqa
In terms of the coherence vector $ \bmrho$ and using the isomorphism \eqref{eq:lie-brack-adj1} (meaning $  {\bm A}\bmrho(t) \simeq    -i [ H_A , \, \rho ] $), from \eqref{eq:adH_Arho0} if $ \bmrho_\mathfrak{k} =0 $, $ {\bm A} \bmrho =0$, i.e., $ \rho =  \varrho_0 \lambda_0 + \rho_\mathfrak{h} $ is a fixed point.
To show the other direction, notice that in \eqref{eq:adH_Arho1} $  {\cal E}_j - {\cal E}_\ell \neq 0 $ $ \forall \;  (j\ell )$ $  1\leqslant j < \ell \leqslant N-1 $, since $ H_A $ is nondegenerate.
Hence whenever $ {\cal F}_\mathfrak{k} ( \rho ) \neq 0$, $ {\bm A} \bmrho = {\bm A} \bmrho_\mathfrak{k} \neq 0$, because of the invariance of the $ \mathfrak{k}_{pq} $ subspaces under $ \mathfrak{h}$, see also \eqref{eq:Cartan-sub-brack1}.
Therefore when $ \bmrho_ \mathfrak{k} \neq 0 $ the unforced system flows along nontrivial periodic orbits.
Condition~\ref{item:drift1} of the last part also follows from \eqref{eq:Cartan-sub-brack1}.
Since $ \bmrho_\mathfrak{h} (t) = \bmrho_\mathfrak{h} (0) $, it must be $ \| \bmrho_\mathfrak{k} (t) \| = {\rm const } $ $ \forall \, t $.
This, together with the invariance property \eqref{eq:Cartan-sub-brack1} yields \ref{item:drift2}.
Finally, Item~\ref{item:drift3} follows from $ \rho_\mathfrak{k} $ never being fixed under the flow of the drift.
To see it, consider a small time increment $ \delta t $.
In the first order approximation, one can write 
\[
\bmrho(t + \delta t ) = e^{ \delta t {\bm A} } \bmrho(t) = \left( I + \delta t {\bm A} \right) \bmrho(t) 
\]
i.e., the increment at $ \delta t $ is given by \eqref{eq:adH_Arho1} and the claim follows from the fact that $ H_A $ is strongly regular, i.e., $ {\cal E}_j - {\cal E}_\ell \neq {\cal E}_p -{\cal E}_q  $, $ ( j\ell ) \neq (pq) $ $ j\neq \ell $, $ p\neq q $.\qed

Since $ {\rm dim} ( {\bm A}) = n > N $, the stationary Liouville equation has more eigenvalues than those referable to the eigenvalues of the corresponding Schr\"{o}dinger equation. 
From $ {\bm A} = -i \ad_{H_A} $, these are the {\em roots} of the Lie algebra $ \mathfrak{su}(N) $ computed at the element $ H_A $ of the Cartan subalgebra $ \mathfrak{h} $, and, from \eqref{eq:adH_Arho1}, they correspond to the transition frequencies $ {\cal E}_j - {\cal E}_\ell $.
The regularity of $ H_A $ guarantees that these extra eigenvalues of $ {\bm A} $ are all nonzero: $ \dim \left( {\rm ker} ({\bm A})\right) = N-1 = \dim (\mathfrak{h} ) $, thus providing an alternative proof of the first part of Proposition~\ref{cor:diag-comm1}.

For $ {\rm tr} ( \rho^2 ) =1$, the mapping $ \ket{\psi} \to \ketbra{\psi}{\psi} $ eliminates the ambiguity in the (unobservable) global phase: $ \ketbra{\psi}{\psi}  = e^{i \varphi }  \ketbra{\psi}{\psi} e^{-i \varphi } $ $ \forall \, \varphi\in \mathbb{R}$ and, as before, the same property holds also for mixed states.
Proposition~\ref{cor:diag-comm1} affirms that, consequently, the corresponding one-parameter orbit passing through each eigenstate $ \ket{\psi_j} $ (due to the global phase) collapses into a fixed point of the unforced Liouville equation. 
Rephrasing in terms of density operators (part of) Proposition~\ref{cor:diag-comm1}, we have the following.
\begin{corollary}
\label{cor:diag-comm2}
Any diagonal density operator is a fixed point of \eqref{eq:Liouville1} when $ u=0$.
More generally, for any density operator both the diagonal part and the trace square norm of the off-diagonal part are integrals of motion of \eqref{eq:Liouville1} when $ u=0$.
\end{corollary}

\section{Structure of the state space: complex flag manifolds}
\label{sec:topol-descr}
It is possible to give a more thorough interpretation of Proposition~\ref{cor:diag-comm1} by studying the structure of the manifold in which $ \rho $ is living, call it $ \mathcal{S}  $. $\mathcal{S}  $ is a connected, simply connected submanifold of $  \mathbb{S}^{n-1}  $ (the $ (n-1)$-dimensional sphere of radius $ \| \bmrho \|$) whose dimension depends on the multiplicities of the eigenvalues of $ \rho$. 
For $ N> 2 $, $\mathcal{S}   \subsetneq \mathbb{S}^{n-1}   $, since the Lie group $ {\rm exp} \left( \ad_{\mathfrak{su}(N)} \right) $ is not acting transitively on the entire $ \mathbb{S}^{n-1}  $.
$  \mathcal{S}   $, instead, is a homogeneous space of $ {\rm exp} ( \ad_{\mathfrak{su}(N)}) $, the action being left matrix multiplication, and can be described as a (co)adjoint orbit of $ SU(N)$ on its Lie algebra as follows.
Consider a diagonal density $  \tilde{\bmrho}_\mathfrak{h} \in  \mathcal{S}  $.
Call $ C_{\tilde{\bmrho}_\mathfrak{h} } $ the stabilizer of $  \tilde{\bmrho}_\mathfrak{h} $, $ C_{\tilde{\bmrho}_\mathfrak{h} } = \{ g \in {\rm exp} \left( \ad_{\mathfrak{su}(N)} \right) \text{ s.t. } g \tilde{\bmrho}_\mathfrak{h} = \tilde{\bmrho}_\mathfrak{h} \} $.
Because of the identification (up to the imaginary unit) of the density operators with (a convex set in) $ \mathfrak{su}(N)$, the coset space $ {\rm exp} \left( \ad_{\mathfrak{su}(N)} \right) /  C_{\tilde{\bmrho}_\mathfrak{h} } $ is the adjoint orbit of $ SU(N)$ on its Lie algebra passing through $ \tilde{\bmrho}_\mathfrak{h}$.
Because of transitivity, this orbit can be identified with $ \mathcal{S}  $: $ \mathcal{S}   = {\rm exp} \left( \ad_{\mathfrak{su}(N)} \right) /  C_{\tilde{\bmrho}_\mathfrak{h} } \, \tilde{\bmrho}_\mathfrak{h} $.
The $\dim (  \mathcal{S}   ) $ is always even (each (co)adjoint orbit has a symplectic structure as is well-known).
The orbit $ \mathcal{S}   $ is transverse to $ \mathfrak{h} $ and meets $ \mathfrak{h} $ in a number of disjoint points equal to the number of distinct permutations of the entries of $  \tilde{\bmrho}_\mathfrak{h} $.
Such number is equal to the cardinality of the Weyl group as well as to the Euler characteristic $ \chi ( \mathcal{S}   )$ of the orbit, see \cite{Ercolessi1,Zyczkowski2} and Theorem E.2 of \cite{Frankel1}.
These points form the vertices of a polygon in the $ N-1 $-dimensional eigenensemble sitting in $ \mathfrak{h}$ and are sometimes denoted Weyl chambers.
Inspired by the $ \mathbb{S}^2 $ case (see Example~\ref{example:2-lev1} below), we shall call them {\em antipodal}.
If $ \tilde{\rho}_\mathfrak{h} = {\rm diag} \left( \tilde{w}_1, \ldots, \tilde{w}_N \right) $, $ \sum_{j=1}^N \tilde{w}_j =1 $, $ 0 \leqslant \tilde{w}_j \leqslant 1 $, then the $ \chi(  \mathcal{S}  ) -1 $ antipodal points are given by $  {\rm diag} \left( \tilde{w}_{\sigma(1)}, \ldots, \tilde{w}_{\sigma(N)} \right) $ with $ \sigma(1), \ldots, \sigma(N) $ a permutation of $ 1, \ldots , N $ such that $   {\rm diag} \left( \tilde{w}_{\sigma(1)}, \ldots, \tilde{w}_{\sigma(N)} \right) \neq \tilde{\rho}_\mathfrak{h}$.
While the topology of the diagonal coset representatives is particularly easy to visualize, the entire orbit enjoys the same topological structure of $ \tilde{\bmrho}_\mathfrak{h}$.
To see it, simply notice that applying a rotation in $ {\rm exp} ( \ad_{\mathfrak{su}(N) } ) $ to two or more diagonal antipodal points they remain antipodal.
Since $ {\rm exp} ( \ad_{\mathfrak{su}(N) } ) $ acts transitively on $ {\cal S}   $, this is true on the entire orbit: each $ \bmrho \in  {\cal S}   $ has $  \chi(  \mathcal{S}    ) -1 $ antipodal states in $ {\cal S}   $.
What is not known a priori is the isotropy subgroup, which depends on $ {\cal J}$.
In fact, the convex set of $N$-level density operators is foliated into leaves of different dimensions, depending on the number of distinct $ w_j$ and on their multiplicities.
For example, for a pure/pseudopure state $ {\cal J} = \{ w_1, w_2, \ldots, w_2 \} $, $w_1 \neq w_2 $, $ w_1+ (N-1)  w_2 =1 $, $  \mathcal{S}   ={\rm exp} ( \ad_{\mathfrak{su}(N)}) /\left( \mathbb{S}^1 \times {\rm exp} ( \ad_{\mathfrak{su}(N-1)})\right)  $ and $ \dim (  \mathcal{S}   ) = 2N-2$.
In the pure state case $ {\cal J} = \{ 1, 0, \ldots, 0 \} $, it is well-known that the map $ \ket{\psi} \to \ketbra{\psi}{\psi} $ can be seen as a Hopf fibration: $ \mathbb{S}^{2N-1} \xrightarrow{\;\mathbb{S}^1 \;} \mathcal{S}  = \mathbb{C}P^{N-1} $, with fibers representing the global phase. 
At the other extreme, if $ {\cal J} = \{ w_1, w_2, \ldots, w_N \} $, $w_j \neq w_\ell $, $ \sum_{j=1}^N w_j =1 $, then $  \mathcal{S}   ={\rm exp} ( \ad_{\mathfrak{su}(N)}) /\left( \mathbb{S}^1 \right)^N $ of dimension $ N^2 -N $.
Hence if $ m= \dim(  \mathcal{S}   ) $, $ 2N-2 \leqslant m \leqslant N^2-N$, $m $ even.
In between lies the flag manifolds with flag determined by the multiplicities of $ w_j$. If such multiplicities are given by $ j_1, \ldots, j_\ell$, $ j_1 + \ldots + j_\ell = N $, $ 2\leqslant \ell \leqslant N $, 
\[   
\mathcal{S}   ={\rm exp} ( \ad_{\mathfrak{su}(N)}) /\left( {\rm exp} ( \ad_{\mathfrak{su}(j_1)}) \times \ldots \times {\rm exp} ( \ad_{\mathfrak{su}(j_\ell)}) \times \left( \mathbb{S}^1 \right)^{\ell-1} \right) .
\]
When $ \ell =2 $ we have Grassmannian manifolds.
Normally, in the literature these are known as {\em complex flag manifolds} and are given directly in terms of unitary group actions, see \cite{Adelman1,Boya1,Zyczkowski2}:  
\beqan
\mathcal{S}  & = & U(N) / \left( U(j_1) \times \ldots \times U(j_\ell )  \right) , \qquad  j_1 + \ldots + j_\ell = N , \qquad 2\leqslant \ell \leqslant N.\\
& = &  SU(N) / \left( SU(j_1) \times \ldots \times SU(j_\ell ) \times \left( \mathbb{S}^1 \right)^{\ell-1}  \right)
\eeqan
In terms of unitary actions, the two extreme cases of pure states and all different eigenvalues are, respectively, $  \mathcal{S}  = U(N) / \left( U(N-1) \times U(1) \right) = SU(N)/ \left( SU(N-1) \times \mathbb{S}^1 \right) $ and $  \mathcal{S}  = U(N) / \left( U(1) \right)^N = SU(N)/ \left( \mathbb{S}^1 \right) ^{N-1}  $.
The description adopted here is just an isomorphic real representation of such complex flag manifolds deriving from the use of the adjoint representation.

\begin{example}  
\label{example:2-lev1}
$ N=2 $, $ {\cal J} = \{ 1, 0 \} $.
The case $N=2$ is the only easy one, as $ {\cal S}  = \mathbb{S}^2 \simeq \mathbb{C} P^1 $.
On the great horizontal circle of $ \mathbb{S}^2  $, $ \bmrho = \bmrho_\mathfrak{k} $.
In terms of the Bloch vector, the diagonal antipodal states become the north and south poles of the Bloch sphere,
\beqan
\rho_{v_1} = \ketbra{0}{0} = {\rm diag} \left( 1, \, 0  \right) & \Longleftrightarrow &\bmrho_{v_1} = \begin{bmatrix} 0 & 0 & \frac{1}{\sqrt{2}} \end{bmatrix}^T \\
\rho_{v_2} = \ketbra{1}{1} = {\rm diag} \left( 0, \, 1  \right) & \Longleftrightarrow &\bmrho_{v_2} = \begin{bmatrix} 0 & 0 & -\frac{1}{\sqrt{2}} \end{bmatrix} ^T 
\eeqan
and $ \mathfrak{h}$, $ {\rm dim} (\mathfrak{h}) =1 $, corresponds to the vertical line passing through $ \bmrho_{v_1}$, $\bmrho_{v_2} $.
Everything extends unchanged to mixed states, since $ {\cal S} $ is still equal to $ \mathbb{S}^2 $ regardless of the purity.
Since each $ {\cal S}  $ crosses $ \mathfrak{h} $ exactly twice, $  \chi ( \mathcal{S}  )  =2 $.
For any $ \bmrho \in \mathcal{S}  $ the antipodal state is $ - \bmrho $. 

\qed

\end{example}
\begin{example}  
$ N=3 $, $ {\cal J} = \{ 1, 0, 0 \} $.
Since the isotropy subgroup in this case is $ SO(3)\times \mathbb{S}^1 $ of dimension 4 (recall that $ \dim(\ad_{\mathfrak{su}(3)} ) = 8$), $ \dim( \mathcal{S}  ) =4$ and $ \chi (  \mathcal{S}  ) =3 $.
Following the standard ordering convention, the 3-level Gell-Mann basis (see e.g. \cite{Georgi1}, p. 99) is
\[ 
\{ \lambda_{\mathfrak{k}, \Re, 12} , \,\lambda_{\mathfrak{k}, \Im, 12} , \, \lambda_{\mathfrak{h}, 1} , \,  \lambda_{\mathfrak{k}, \Re, 13} , \,\lambda_{\mathfrak{k}, \Im, 13} , \, \lambda_{\mathfrak{k}, \Re, 23} , \,\lambda_{\mathfrak{k}, \Im, 23} , \,\lambda_{\mathfrak{h}, 2} \}.
\] 
The three diagonal antipodal states are 
\beqan
\rho_{v_1} = {\rm diag} \left( 1, \, 0, \, 0 \right) & \Longleftrightarrow &\bmrho_{v_1} = \begin{bmatrix} 0 & 0 & \frac{1}{\sqrt{2}} & 0 & 0 & 0 & 0 & \frac{1}{\sqrt{6}} \end{bmatrix}^T  \\
\rho_{v_2} = {\rm diag} \left( 0, \, 1, \, 0 \right) & \Longleftrightarrow &\bmrho_{v_2} = \begin{bmatrix} 0 & 0 & -\frac{1}{\sqrt{2}} & 0 & 0 & 0 & 0 & \frac{1}{\sqrt{6}} \end{bmatrix}^T  \\
\rho_{v_3} = {\rm diag} \left( 0, \, 0, \, 1 \right) & \Longleftrightarrow &\bmrho_{v_3} = \begin{bmatrix} 0 & 0 & 0 & 0 & 0 & 0 & 0 & -\frac{2}{\sqrt{6}} \end{bmatrix} ^T .
\eeqan
The structure of $ {\cal S} \subset \mathbb{S}^7 $ is studied in detail in \cite{Ercolessi1,Chaturvedi1,Kimura1,Schirmer7}.
In terms of the coherence vector, one has that each state of a triplet of antipodal states is at an angle of $ \frac{2\pi}{3} $ from the other two, see Fig.~\ref{FIG3lev-ref2}.
In particular, if $ \bmrho \in \mathcal{S}  $, then $ -\bmrho \notin {\cal S}  $: the single antipodal point of the case $N=2 $ is replaced by two symmetrically distributed and equidistant antipodal points.
As expected, only $ \lambda_{\mathfrak{h},1} $ and $ \lambda_{\mathfrak{h},2} $ are of concern when $ \rho $ is diagonal.  
In an attempt to visualize the entire $ {\cal S}$ manifold, one should replace each edge connecting to vertices with a sphere $ \mathbb{S}^2 $.
The same numbers occur for pseudopure states, which have the same twofold degeneracy and are obtained by rescaling down the coherence vector $ \bmrho_{v_j} $ by a constant factor.
For the all different eigenvalue case $ \rho_\mathfrak{h} = {\rm diag} \left( w_1, \, w_2, \, w_3 \right) $, $ w_j \neq w_k $, which is the generic case, the stabilizer is the torus $ \mathbb{S}^1 \times \mathbb{S}^1  $, $\mathcal{S} ={\rm exp} ( \ad_{\mathfrak{su}(N)}) / (\mathbb{S}^1 \times \mathbb{S}^1)  $ and $ \dim( \mathcal{S}  ) =6$.
The only diagonal matrices that are conjugate with $ \rho_\mathfrak{h} $ are its five element permutations, i.e., $ \chi ( \mathcal{S}  ) =6 $ in this case.
The six vertices are given by 
\beq
\begin{split}
{\rm diag} \left( a, \, b, \, c \right)  \Longleftrightarrow &\bmrho_{g_1} = \begin{bmatrix} 0 & 0 & \frac{a-b}{\sqrt{2}} & 0 & 0 & 0 & 0 & \frac{a+b-2c}{\sqrt{6}} \end{bmatrix}^T  \\
{\rm diag} \left( b, \, a, \, c \right)  \Longleftrightarrow &\bmrho_{g_2} = \begin{bmatrix} 0 & 0 & \frac{b-a}{\sqrt{2}} & 0 & 0 & 0 & 0 & \frac{a+b-2c}{\sqrt{6}} \end{bmatrix}^T  \\
{\rm diag} \left( c, \, a, \, b \right)  \Longleftrightarrow &\bmrho_{g_3} = \begin{bmatrix} 0 & 0 & \frac{c-a}{\sqrt{2}} & 0 & 0 & 0 & 0 & \frac{a+c-2b}{\sqrt{6}} \end{bmatrix}^T  \\
{\rm diag} \left( c, \, b, \, a \right)  \Longleftrightarrow &\bmrho_{g_4} = \begin{bmatrix} 0 & 0 & \frac{c-b}{\sqrt{2}} & 0 & 0 & 0 & 0 & \frac{b+c-2a}{\sqrt{6}} \end{bmatrix}^T  \\
{\rm diag} \left( b, \, c, \, a \right)  \Longleftrightarrow &\bmrho_{g_5} = \begin{bmatrix} 0 & 0 & \frac{b-c}{\sqrt{2}} & 0 & 0 & 0 & 0 & \frac{b+c-2a}{\sqrt{6}} \end{bmatrix}^T  \\
{\rm diag} \left( a, \, c, \, b \right)  \Longleftrightarrow &\bmrho_{g_6} = \begin{bmatrix} 0 & 0 & \frac{a-c}{\sqrt{2}} & 0 & 0 & 0 & 0 & \frac{a+c-2b}{\sqrt{6}} \end{bmatrix}^T .
\end{split}
\label{eq:vertex-generic-3lev}
\eeq
The 6 vertices correspond to $ \frac{2\pi}{3} $ rotations, plus their reflections around the three axes bisecting the triangle of pure states, see Fig.~\ref{FIG3lev-ref2}.
Represented in the $ ( \lambda_{\mathfrak{h},1}, \, \lambda_{\mathfrak{h},2} ) $ plane, the eigenensemble of $ \rho $ has the shape of a hexagon inscribed in the triangle having as vertices the pure states, see Fig.~\ref{FIG3lev-ref2}.
\begin{figure}[ht]
\begin{center}
\includegraphics[width=5cm]{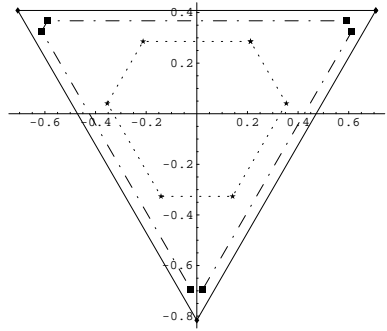}
 \caption{Vertices of the eigenensemble in the $ ( \lambda_{\mathfrak{h},1}, \, \lambda_{\mathfrak{h},2} ) $ plane for $ N =3$. Vertices joined by the solid line: pure states. Vertices joined by the dotted line: ``generic'' mixed state with all different eigenvalues: $a=3/5$, $b=3/10$ and $ c=1/10$. Vertices joined by the dashed-dotted line: another ``generic'' mixed state with eigenvalues $a=9/10$, 
$b=2/30$ and $ c=1/30$. }
\label{FIG3lev-ref2}
\end{center}
\end{figure}

\qed

\end{example}

\section{Feedback stabilization for $N$-level quantum ensembles}
\subsection{Problem formulation}

For the system \eqref{eq:Liouv-vect-coh1}, we are interested in the problem of tracking a periodic orbit. 
More precisely, the stabilization problem is the following.

\begin{quote}
{\em Given $ \rho_d \in {\cal S}   $, find $ u = u (\bmrho_d, \bmrho ) $ such that, for $ t\to \infty $, $ \bmrho  \to \bmrho_{d} $, where 
\beq
\dot{\bmrho}_d = {\bm A}_d \bmrho_d,  \qquad {\bm A}_d = {\bm h}_{A_d} \cdot \ad_{\bmlambda_\mathfrak{h}}. 
\label{eq:ref-traj}
\eeq
}
\end{quote}
This is a full state tracking problem which, from Proposition~\ref{cor:diag-comm1}, reduces to stabilization to an equilibrium point when $ \rho_d = \varrho_0 \lambda_0 + \bmrho_{d,\mathfrak{h}} \cdot \bmlambda_\mathfrak{h}$.

\subsection{A modified Jurjevic-Quinn condition and antipodal points}

The algorithm for the feedback design resembles the one used for $ \ket{\psi}$ discussed in \cite{Ferrante1,Vettori1,Grivopoulos1,Mirrahimi1,Mirrahimi2} and indeed the standard Jurdjevic-Quinn method for bilinear systems \cite{Jurdjevic5}.
It consists in choosing a distance-like candidate Lyapunov function $V = V( \bmrho_d, \bmrho) $.
From \eqref{eq:dist-cohvec}, consider \footnote{Notice that rather than the ``distance inherited from the $ \mathbb{S}^{n-1}  $ sphere'' $V$ used here, one could use the $ \mathbb{R}^n $ distance $ \| \delta \bmrho \|^2 = \| \bmrho_d - \bmrho \|^2 $ as candidate Lyapunov function.
Up to a scalar factor, the two give the same gradient, hence the same control design.} $ V =  \| \bmrho \|^2 - \llangle \bmrho_d, \bmrho \rrangle $. If $ \bmrho_d $ obeys to \eqref{eq:ref-traj}, 
\beq
\begin{split}
\dot{V} = \dot{V}( \bmrho_d, \bmrho)  & =  - \llangle {\bm A}_d \bmrho_d, \bmrho \rrangle -  \llangle \bmrho_d, {\bm A} \bmrho  \rrangle - u \llangle \bmrho_d,{\bm B}\bmrho  \rrangle \\
& =   \llangle \bmrho_d,  ( {\bm A}_d-  {\bm A} )  \bmrho  \rrangle - u \llangle \bmrho_d,{\bm B}\bmrho  \rrangle .
\end{split}
\label{eq:dotV-N-lev}
\eeq
This expression has a drift term which is in general sign indefinite and for which there is no global smooth feedback compensation.
Whenever $ \dot{V} $ can be rendered homogeneous in $u$ (namely when $ H_{A_d} = H_A $, which is assumed thereafter), there is an obvious choice of feedback guaranteeing positive semidefiniteness.
For example, 
\beq 
u=  \llangle \bmrho_d,{\bm B}\bmrho  \rrangle 
\label{eq:feed-N-lev1} 
\eeq
makes $ \dot{V} = - u^2 \leqslant 0 $.
In order to apply LaSalle invariance principle, one can try to adapt the ``$\ad$-condition'' of Jurdjevic-Quinn \cite{Jurdjevic5} to the case at hand.
The largest invariant set $ {\cal E} $ in $ {\cal N}=\{ \bmrho \text{ s.t. } \dot{V} =0 \} $ can be computed imposing $ u =\de{u}{t} = \ldots = \frac{ d^\ell u }{d t^\ell} =0 $, where, in correspondence of $ u=0$: 
\[
\de{u}{t} = - \bmrho^T_d \left[ {\bm A} , \,{\bm B}\right] \bmrho =0
\]
and, similarly,
\beq
\de{^\ell u}{t^\ell} = ( -1)^\ell \bmrho^T_d \underbrace{[ {\bm A} , \ldots ,  [ {\bm A} ,}_{\text{ $\ell$ times}}  \,{\bm B}] \ldots ]  \bmrho =0.
\label{eq:deru-k-times}
\eeq
For $ \ket{\psi} $ which is an eigenfunction, this condition, used implicitly in \cite{Vettori1}, is made explicit in \cite{Mirrahimi1} as a Kalman rank condition on the linearized tangent system.
It can be reformulated for densities as follows.
For a given $ \bmrho_d \in {\cal S}   $, write $ \bmrho = \bmrho_d + \delta \bmrho $. 
The linearization of \eqref{eq:Liouv-vect-coh1} around $ \bmrho_d $ 
\beq
\de{\delta \bmrho }{t} =  {\bm A} \delta \bmrho + {\bm b} u
\label{eq:Liouv-lineariz}
\eeq
where 
\beq 
{\bm b} ={\bm B}\bmrho_d ,
\label{eq:b-lineariz1}
\eeq 
is living on $ T_{\bmrho_d} {\cal S}   $.
Since $ \dim ( T_{\bmrho_d} {\cal S}   ) =m $, if the Kalman rank condition
\beq
{\rm rank } \left[ {\bm  b} \,  {\bm A}{\bm  b } \ldots  {\bm A}^{m-1} {\bm  b} \right] = m 
\label{eq:Kalman1}
\eeq
is satisfied, then in the same spirit of the original Jurdjevic-Quinn work, this implies that $ {\cal E}  $ contains no other trajectory than $ \bmrho_d $, at least locally.
Of course if $ \rho_d = \varrho_0 \lambda_0 + \bmrho_{d,\mathfrak{h}} \cdot {\bm \lambda} $, then under \eqref{eq:ref-traj} $ \rho_d (t) = \rho_d (0) $ and ${\bm  b} (t) = {\bm b }(0)$ in \eqref{eq:b-lineariz1}.
However, when $ \rho_d(0) $ is not diagonal, $ \rho_d (t) $ is time-varying and so is $ {\bm b} (t) $, complicating the verification of \eqref{eq:Kalman1} considerably.
Furthermore, in this case the condition \eqref{eq:Kalman1} does not have a global character because of the topological structure of $ {\mathcal{S}}   $.
To see this, consider the case of diagonal density operators in which the linearization \eqref{eq:Liouv-lineariz} is time-invariant.
Call $ \bmrho_p $ an antipodal state of $ \bmrho_d $.
Then also $ \bmrho_p $ is diagonal and so is $ \delta \bmrho_p = \bmrho_p - \bmrho_d $.
Hence $ {\bm A} \delta \bmrho_p =0 $, i.e., \eqref{eq:Liouv-lineariz} has vanishing drift in correspondence of the $ \chi({\mathcal{S}}   ) -1 $ antipodal states of $ \bmrho_d $.
Therefore, in spite of the Kalman condition \eqref{eq:Kalman1}, in this case when checking LaSalle invariance principle $ \left. \de{\delta \bmrho}{t} \right|_{{\cal N}, \bmrho=\bmrho_p } =0 $ for \eqref{eq:Liouv-lineariz} since $ u=0 $ in $ {\cal N}$.
This argument can be generalized as follows.

\begin{proposition}
\label{prop:equil-antipod}
Given $ \bmrho_d \in \mathcal{S}   $, any of the other $ \chi({\cal S}   ) -1$ antipodal states $ \bmrho_p \in \mathcal{S}   $ is an equilibrium point for the system \eqref{eq:Liouv-vect-coh1} with the feedback \eqref{eq:feed-N-lev1}.
\end{proposition}
\proof
We make use of the isomorphism \eqref{eq:lie-brack-adj1}.
For any given $ \rho_p$, $ \rho_d $ diagonal, $ \left[ -i H_B , \rho_p \right] \in \mathfrak{k}/i$ is off-diagonal i.e., ${\bm B} \bmrho_p = \bmrho'_{p,\mathfrak{k}}$.
But if $ \rho_d $ diagonal $ \rho_d = \varrho_0 \lambda_0 + \bmrho_{\mathfrak{h} _d} \cdot {\bm \lambda} $ then $ u = \llangle \bmrho_d ,\,{\bm B}\bmrho_p \rrangle = 
\llangle \bmrho_d ,\,\bmrho'_{p,\mathfrak{k}}\rrangle = 0 $, since $ \mathfrak{h} \perp \mathfrak{k}$.
Hence no feedback is produced.
When instead $ \bmrho_p, \, \bmrho_d \in {\mathcal{S}}   $ are antipodal but not diagonal, then (by construction) $ \exists $ $ {\bm R} \in {\rm exp}( \ad_{\mathfrak{su}(N)} ) $ such that $ \tilde{\bmrho}_p = {\bm R} \bmrho_p $ and $ \tilde{\bmrho}_d = {\bm R} \bmrho_d $ are both diagonal. 
The skew symmetric matrix $ {\bm R}^T ( -i \ad_{H_B})  {\bm R} $ belongs to the adjoint orbit of $ {\rm exp}( \ad_{\mathfrak{su}(N)} ) $ in $  \ad_{\mathfrak{su}(N)} $, hence $ \exists $ $C \in  \mathfrak{su}(N) $ such that $ {\bm R}^T {\bm B} {\bm R}  = \ad_C $.
Therefore $ u =  \llangle \bmrho_d ,\,{\bm B} \bmrho_p \rrangle = \llangle \tilde{\bmrho}_d ,\,{\bm R}^T {\bm B} {\bm R}  \tilde{\bmrho}_p  \rrangle = \llangle \tilde{\bmrho}_d , \ad_C \tilde{\bmrho}_p  \rrangle = 0 $ because $ \ad_C \tilde{\bmrho}_p \simeq [ C, \, \tilde{\rho}_p ] \in \mathfrak{k}/i $, while $ \tilde{\rho}_d \in \mathfrak{h}$.
\qed

\begin{corollary}
\label{cor:equil-antipod-max}
For pure or pseudopure states, the $N$ antipodal points of $ {\cal S} $ are all equidistant. For pure states they are also maximally distant.
\end{corollary}
\proof
It is enough to notice that for any triple of antipodal points $ \rho_{p_1} $, $ \rho_{p_2} $ and $ \rho_{p_3}$ ($ \rho_d $ included), $ \tr{ \rho_{p_1} \rho_{p_2} } = \tr{ \rho_{p_1} \rho_{p_3} } = \tr{ \rho_{p_2} \rho_{p_3} } $, hence $ V( \bmrho_{p_j} , \bmrho_{p_\ell} ) = {\rm const} $ $ \forall \; j, \ell =1, \ldots, N$, $ j\neq \ell$. For pure states, in addition, $ \tr{ \rho_{p_j} \rho_{p_\ell} } = 0$, hence $ V( \bmrho_{p_j} , \bmrho_{p_\ell} ) = {\rm tr} (\rho_{p_j} ^2 ) $ are maximally distant.
\qed 

As will be shown in next Section, the antipodal points are not the only states lacking attractivity, and the linearization alone is not enough to investigate the domain of attraction of the feedback stabilizer.

\begin{remark}
\label{rem:rot-fr}
The trajectory tracking problem presented above admits a reformulation as a point stabilization for a nonautonomous system.
Consider a frame rotating with $ {\bm A}$.
Call $ \widehat{\bmrho}_d $ and $ \widehat{\bmrho} $ the new reference and state vectors.
Then $  \widehat{\bmrho} (t) = e^{ -t {\bm A}} \bmrho (t) $ and $ \widehat{\bmrho}_d (t) = e^{ - t {\bm A} } \bmrho_d (t) = \bmrho_d(0) $, i.e., the reference trajectory becomes a {\em fixed point}.
Using a variation of constants formula, we obtain for \eqref{eq:Liouv-vect-coh1}
\beq
\begin{cases} \dot{\widehat{\bmrho}}  & = u \,e^{ - t {\bm A} }  {\bm B} e^{  t {\bm A} } \widehat{\bmrho} \\
\widehat{\bmrho} (0)  & = \bmrho(0) .
\end{cases}
\label{eq:rot-fr1}
\eeq
The Lyapunov distance is $ V  = \| \widehat{\bmrho}_d ^2 \| - \llangle \widehat{\bmrho}_d, \widehat{\bmrho} \rrangle $ and its derivative $ \dot{V} = u \llangle \widehat{\bmrho}_d, \dot{\widehat{\bmrho}} \rrangle $. The uniformity of the asymptotic stability for the nonautonomous system \eqref{eq:rot-fr1} with the same feedback stabilizer as \eqref{eq:feed-N-lev1} follows directly.
\end{remark}

\subsection{Time-independent convergence conditions}

To formulate a convergence condition in a more geometric manner, rewrite \eqref{eq:deru-k-times} in terms of bilinear forms of skew-symmetric matrices as follows.
Call 
\[
 \mathfrak{W}^\alpha = {\rm span} \left\{ {\bm B}, \, \left[ {\bm A} , \,{\bm B}\right], \ldots , \underbrace{ [ {\bm A} , \ldots ,  [ {\bm A} ,}_{\text{ $ \alpha $ times }}  \,{\bm B}] \ldots ] \right\} 
\] 
and $ \mathfrak{W}_A^\alpha = {\rm span} \left\{{\bm A}, \,\mathfrak{W}^\alpha \right\} $.
$ \mathfrak{W}^\alpha$ contains skew-symmetric matrices, and the conditions $ \de{^\ell u }{t^\ell} $, $ \ell =0, \ldots, \alpha $, written compactly as $ \bmrho_d^T\mathfrak{W}^\alpha  \bmrho $, are bilinear forms.
If $ \bmrho /\!/ \bmrho_d $ then $ \bmrho_d ^T \mathfrak{W}^\alpha \bmrho=0$.
On a sphere $  \bmrho /\!/ \bmrho_d $ means $ \bmrho = \pm \bmrho_d $.
However $ {\cal S}   $ is only a submanifold of $ \mathbb{S}^{n-1}  $, and $ - \bmrho_d $ may not belong to it at all.
$ \mathfrak{W}^\alpha $ is invariant to the so-called ``$ \ad$-brackets'' but not necessarily a Lie subalgebra.
Of course if for the system \eqref{eq:Liouv-vect-coh1} the $ \ad$-brackets are generating, i.e., if, for some $ \alpha $, $  \mathfrak{W}_A^\alpha= \ad_{\mathfrak{su}(N)} $, then almost global convergence on $ {\cal S} $ is always verified.
However we have the following negative result.
\begin{lemma}
\label{prop:ad-never-N}
If $ N \geqslant 3 $, for the system \eqref{eq:Liouv-vect-coh1} with $ H_A $ strongly regular and $ {\rm Graph} (H_B) $ connected, $ {\rm Lie} ( \mathfrak{W}_A^\alpha ) = \ad_{\mathfrak{su}(N)} $ but $  \mathfrak{W}_A^\alpha\subsetneq  \ad_{\mathfrak{su}(N)} $ $ \forall \, \alpha >0 $.
\end{lemma}
\proof
Since $ H_A $ is strongly regular and $ {\rm Graph } (H_B ) $ is connected, it follows from Theorem~2 of \cite{Cla-contr-root1} that the smallest subalgebra containing $ -i H_A $, $ -i H_B $ is $ \mathfrak{su}(N)$. Hence the same holds for the adjoint representation.
For the second part, recall that $ \dim(\mathfrak{h} ) = N-1 $.
From the Lie bracket relations \eqref{eq:Cartan-sub-brack1}, $ {\bm A} \in \ad_\mathfrak{h} $, ${\bm B}\in \ad_\mathfrak{k} $ implies $ \left[ {\bm A} , \ldots ,  \left[ {\bm A} , \, {\bm B} \right] \ldots \right] \in \ad_\mathfrak{k} $.
Even adding $ {\bm A} $, $ \mathfrak{W}_A^\alpha $ alone cannot fully generate $ \ad_\mathfrak{h} $ for any $ \alpha $.
\qed

The first part of Lemma~\ref{prop:ad-never-N} is also known as the strong accessibility condition \cite{Nijmeijer1}. Since $ \ad_{\mathfrak{su}(N)} $ is compact, it suffices for controllability. 
The second part is the Jurdjevic-Quinn condition mentioned above.
If $ H_B $ is not off-diagonal, then the statement of Lemma~\ref{prop:ad-never-N} should be reformulated as ``$ N>3 $''.


The following Theorem provides a time-independent condition for asymptotic stabilizability to any $ \rho_d \in {\cal S} $, and a global description of the region of attraction of the controller.

\begin{theorem}
\label{thm:conver-trajtr}
Consider the system \eqref{eq:Liouv-vect-coh1} with the feedback \eqref{eq:feed-N-lev1}, where $ \rho_d \in {\cal S}  $ obeys to \eqref{eq:ref-traj}. 
Assume that $ H_A $ is strongly regular and that $ H_B $ is such that $ ( h_{B, \Re, j\, j+1},  h_{B, \Im, j\, j+1}) \neq ( 0, \, 0)$.
An initial condition $ \rho(0)\in {\cal S} $ is asymptotically converging to $ \rho_d(t)$ if
\begin{enumerate}
\item \label{Ass0-thm1} $ \rho(0) $ is not an antipodal point of $ \rho_d (0)$,
\item \label{Ass1-thm1} $ {\cal F} \left([ H_B, \, \rho_d ] \right) \cap  {\cal F} \left( \rho (0) \right) \neq 0$,
\item \label{Ass2-thm1} $ {\rm Card} {\cal F}_\mathfrak{k} \left([ H_B, \, \rho_d ] \right) \geqslant m/2 $
\end{enumerate}
where $ {\rm Card} $ denotes the number of pairs of indexes in $ {\cal F}_\mathfrak{k}$.
\end{theorem}

In order to prove the Theorem we need a few preliminary results.

\begin{lemma} 
\label{lem:Kalm-equiv}
Under the assumption of strong regularity of $ H_A $, the following three conditions are equivalent:
\begin{enumerate}
\item \label{Ass1-lemm-Kalm} the Kalman rank condition \eqref{eq:Kalman1} is satisfied;
\item \label{Ass2-lemm-Kalm} $ {\rm rank} \left( \mathfrak{W}^{m-1} \bmrho_d \right) =m$;
\item \label{Ass3-lemm-Kalm} $ {\rm Card} {\cal F}_\mathfrak{k}  ( [H_B, \rho_d ] ) \geqslant m/2 $.
\end{enumerate}
\end{lemma}

\proof
Given $ C\in \mathfrak{su}(N)$, strong regularity of $ H_A $ implies that $ C , [H_A , C ] , \ldots, [H_A, \ldots , [H_A, C] \ldots ] $ are all linearly independent up to a number $ \alpha -1 $ , $ \alpha = 2\, {\rm Card } {\cal F}_\mathfrak{k} (C) $, of nested $ H_A $ commutators, see Theorem~2 in \cite{Cla-contr-root1}.
Using $ C = {\bm c } \cdot {\bm \lambda } $ and the isomorphism given by the adjoint representation, the vectors $ {\bm c }, {\bm A} {\bm c }, \ldots  {\bm A}^{\alpha-1}  {\bm c} $ are all linearly independent.
If $ {\bm c } = {\bm b} = {\bm B}\bmrho_d $ as in \eqref{eq:b-lineariz1}, then this is the Kalman controllability condition provided $ \alpha \geq m $.
If $ {\bm b} = {\bm b}_\mathfrak{h} + {\bm b}_\mathfrak{k} $, $  {\bm A}  {\bm b}_\mathfrak{h} =0 $, hence only the off-diagonal part of $ [ H_B , \rho_d ] $ matters. 
The support $ \mathfrak{f}_\mathfrak{k} ([ H_B, \rho_d]) $ intersects a number of ``root spaces'' $ \mathfrak{k}_{j\ell} $ (each has real dimension 2) equal to $ {\rm Card} {\cal F}_\mathfrak{k} ([ H_B, \rho_d]) $.
Furthermore, since $ -i H_A \in \mathfrak{su}(N) $, the invariance property \eqref{eq:Cartan-sub-brack1} applies.
Written in terms of the original commutators 
\beq 
{\cal F}_\mathfrak{k} ([ H_B, \rho_d]) =  {\cal F}_\mathfrak{k} ([ H_A , [ H_B, \rho_d]] ) = \ldots =  {\cal F}_\mathfrak{k} ([ H_A , \ldots, [H_A , [ H_B, \rho_d]]\ldots ] ) .
\label{eq:supp-comm-seq}
\eeq
For the $ \ell$-th order commutator of $ \mathfrak{W}^{\alpha-1}$, one has the binomial-like expansion:
\beq
\begin{split}
&  \underbrace{ [  {\bm A}, \ldots , [  {\bm A} }_{\text{$ \ell$ times}} ,{\bm B}]\ldots ] \\
& \qquad =  {\bm A}  ^\ell   {\bm B}
+ (-1)^1 {\ell \choose 1}  {\bm A}  ^{\ell-1}  {\bm B}   {\bm A}  + \ldots \\
& \qquad \; + (-1)^{\ell-1} {\ell \choose \ell - 1}  {\bm A}    {\bm B}  {\bm A} ^{\ell-1} 
+  (-1)^\ell  {\bm B}  {\bm A} ^\ell.
\end{split}
\label{eq:WequalKalman}
\eeq
After the linearization around $ \rho_d $, only the first term of this expression is retained.
Since $ -i H_A \in \mathfrak{h}$, from \eqref{eq:Cartan-sub-brack1}, $ \mathfrak{f}_\mathfrak{k} \left(  {\bm A}  ^\ell \bmrho_d \right)  =  \mathfrak{f}_\mathfrak{k} ( \bmrho_d) $, while $ \mathfrak{f}_\mathfrak{h} \left( {\bm A} ^\ell \bmrho_d \right)  = 0 $, $ \forall \, \ell \geqslant 1 $.
From \eqref{eq:supp-comm-seq}, $ {\cal F}_\mathfrak{k}$ is the same for all terms in \eqref{eq:WequalKalman}, and similarly,
\beq
{\cal F}_{\mathfrak{k}} \left( {\bm B}  \bmrho_d \right) = {\cal F}_{\mathfrak{k}} \left( [  {\bm A} , {\bm B}] \bmrho_d \right)= \ldots = {\cal F}_{\mathfrak{k}} \left(  [  {\bm A}, \ldots , [  {\bm A}  ,{\bm B}]\ldots ]  \bmrho_d \right) .
\label{eq:supp-ad-comm}
\eeq
In summary, strong regularity of $ H_A $ guarantees the full spanning of a linear space whose dimension is determined uniquely by $ {\rm Card} {\cal F}_\mathfrak{k} ([ H_B, \rho_d]) $. This space is identifiable with the tangent space $ T_{\bmrho_d} {\cal S}   $ as well as with $ \mathfrak{W}^{m-1} \bmrho_d $. The equivalence of the three conditions follows consequently.
\qed

\begin{remark}
Lemma~\ref{prop:ad-never-N} and Condition~\ref{Ass2-lemm-Kalm} of Lemma~\ref{lem:Kalm-equiv} imply that although the vector space $ \mathfrak{W}^{m-1} $ is never the entire Lie algebra $ \ad_{\mathfrak{su}(N)} $ acting transitively on $ {\cal S}$, it may nevertheless span the entire tangent space at a point.
The same holds for the Kalman controllability.
\end{remark}

\begin{remark} In general $  {\rm Card} {\cal F}_\mathfrak{k} ( H_B )\neq  {\rm Card} {\cal F}_\mathfrak{k} ([ H_B, \rho_d]) $, hence the controllability of the linearization depends from the reference trajectory $ \rho_d $ chosen.
The meaning of Lemma~\ref{lem:Kalm-equiv} is that in order to have linear controllability the reference trajectory \eqref{eq:ref-traj} must be ``rich enough'' along $ H_B $.
\end{remark}

\begin{remark}
While the Kalman condition \eqref{eq:Kalman1} seems time-varying as soon as $ \rho_{d,\mathfrak{k}} \neq 0 $, the equivalent condition~\ref{Ass3-lemm-Kalm} of Lemma~\ref{lem:Kalm-equiv} is always time-independent since $ {\cal F} \left([ H_B, \, \rho_d(0) ] \right) =  {\cal F} \left([ H_B, \, \rho_d(t) ] \right) $.
\end{remark}

\begin{remark}
The conditions of Lemma~\ref{lem:Kalm-equiv} depend on $ \rho_d$, $ H_A $ and $H_B $ but not on the state $ \rho$, meaning that alone they are not enough to guarantee convergence of a given $ \rho(0)$.
\end{remark}

The Lyapunov derivative in \eqref{eq:dotV-N-lev} is made homogeneous in $ u$ by the cancellation of the drift term and therefore the notion of attractivity provided by $ \dot{V}$ must be rendered invariant under such flow (in a way similarly to the orbital stabilization problem, see \cite{Bacciotti1}).
The following Lemma gives an alternative attractivity condition which is fully invariant under the drift and generically (i.e., almost always under $ e^{ t {\bm A}} $) equivalent to the usual Lyapunov convergence property.
This last in fact may fail in isolated points: certain critical points of $ V$ are not invariant under the flow of the drift (see Section~\ref{sec:cases}).

\begin{lemma} 
\label{lem:support-equiv}
Consider the system \eqref{eq:Liouv-vect-coh1} with the feedback \eqref{eq:feed-N-lev1}, where $ \rho_d $ obeys to \eqref{eq:ref-traj}. 
If $ H_A $ is strongly regular, the following conditions are generically equivalent under the flow of the drift term:
\begin{enumerate}
\item \label{en:item1-supp-equiv} $ {\cal F} \left([ H_B, \, \rho_d ] \right) \cap  {\cal F} \left( \rho  \right) \neq 0$;
\item \label{en:item2-supp-equiv} $ \dot V ( \bmrho_d, \bmrho) < 0 $;
\item \label{en:item3-supp-equiv} $ \bmrho_d^T \mathfrak{W}^{\alpha} \bmrho = \{ z_0, z_1, \ldots z_{\alpha} \} $, $ z_j \neq 0$.
\end{enumerate}
\end{lemma}

\proof
Clearly $ \dot V = - \llangle \bmrho_d ,{\bm B}\bmrho \rrangle^2  < 0 $ implies $ {\cal F} \left([ H_B, \, \rho_d ] \right) \cap  {\cal F} \left( \rho \right) \neq 0$.
To prove that also the contrary is generically true, it is enough to show that when $ {\cal F} \left([ H_B, \, \rho_d ] \right) \cap  {\cal F} \left( \rho  \right) \neq 0$ the zero crossing of the inner product can occur only at isolated points along the trajectories of the closed loop system.
Assume Item~\ref{en:item1-supp-equiv} holds and, at time $ t$, $ \llangle \bmrho_d ,{\bm B}\bmrho \rrangle =0$.
If $ \delta t $ is a small time increment, then from Item~\ref{item:drift1} of Proposition~\ref{cor:diag-comm1}, $  {\cal F} \left([ H_B, \, \rho_d ] \right)$ and $ {\cal F} \left( \rho  \right) $ remain the same, while, from Item~\ref{item:drift3} of Proposition~\ref{cor:diag-comm1} 
\beqan 
( \bmrho_{\mathfrak{k}, \Re, j\ell } (t + \delta t ), \bmrho_{\mathfrak{k}, \Im, j\ell }(t + \delta t ) )  & \neq & ( \bmrho_{\mathfrak{k}, \Re, j\ell } (t),\bmrho_{\mathfrak{k}, \Im, j\ell }(t  ))   \\
\bmrho_{\mathfrak{h}, j }(t + \delta t ) & = & \bmrho_{\mathfrak{h}, j }(t  ) \\( \bmrho_{d,\mathfrak{k}, \Re, j\ell } (t + \delta t ),\bmrho_{d,\mathfrak{k}, \Im, j\ell }(t + \delta t ) )  & \neq & ( \bmrho_{d,\mathfrak{k}, \Re, j\ell } (t ), \bmrho_{d,\mathfrak{k}, \Im, j\ell }(t  ))   \\
\bmrho_{d,\mathfrak{h}, j }(t + \delta t ) & = & \bmrho_{d,\mathfrak{h}, j }(t) .
\eeqan
If $  {\cal F}_\mathfrak{h} \left([ H_B, \, \rho_d ] \right) \cap  {\cal F}_\mathfrak{h} \left( \rho  \right) \neq 0$, then from the last row of \eqref{eq:Cartan-sub-brack2} only $ \bmrho_{d,\mathfrak{k}} $ matters in the computation of $ {\cal F}_\mathfrak{h} \left( [ H_B, \rho_d] \right) $, and $  \llangle \bmrho_{d,\mathfrak{k}} (t + \delta t ) ,{\bm B}\bmrho_\mathfrak{h} (t + \delta t ) \rrangle  \neq 0 $ since $ \bmrho_{d,\mathfrak{k}}  (t + \delta t )  \neq  \bmrho_{d,\mathfrak{k}}  (t  ) $, while $ \bmrho_\mathfrak{h}  (t + \delta t )  =  \bmrho_\mathfrak{h}  (t  ) $.
If, instead, $  {\cal F}_\mathfrak{k} \left([ H_B, \, \rho_d ] \right) \cap  {\cal F}_\mathfrak{k} \left( \rho  \right) \neq 0$, then we have two possible contributions to consider: $ {\cal F}_\mathfrak{k} \left([ H_B, \, \rho_{d,\mathfrak{h}} ] \right) $ and $ {\cal F}_\mathfrak{k} \left([ H_B, \, \rho_{d,\mathfrak{k}} ] \right) $.
In the first case the conclusion follows from the same argument used above since now $ \bmrho_{d,\mathfrak{h}}  (t + \delta t )  =  \bmrho_{d,\mathfrak{h}}  (t  ) $ while $ \bmrho_\mathfrak{k}  (t + \delta t )  \neq  \bmrho_\mathfrak{k}  (t  ) $.
In the second case it follows from the observation that $  {\cal F}_\mathfrak{k} \left([ H_B, \, \rho_{d,\mathfrak{k}} ] \right) \cap  {\cal F}_\mathfrak{k} \left( \rho  \right) \neq 0$ implies $  {\cal F}_\mathfrak{k} (\rho_{d,\mathfrak{k}} ) \neq  {\cal F}_\mathfrak{k} \left( \rho  \right) $ (see the explicit computations of the commutators in \eqref{eq:comm-kk}).
The general case $ {\cal F} \left([ H_B, \, \rho_d ] \right) \cap  {\cal F} \left( \rho  \right) \neq 0$ is the sum of the two situations just described.
Concerning Item~\ref{en:item3-supp-equiv}, it is enough to notice that generically $  \llangle \bmrho_d ,{\bm B}\bmrho \rrangle  \neq 0$ if and only if $  \llangle \bmrho_d , [ {\bm A}, \ldots, [ {\bm A},{\bm B}]\ldots ]  \bmrho \rrangle \neq 0$.
The argument is of the same type used in the proof of Lemma~\ref{lem:Kalm-equiv}. 
For example if $  {\cal F}_\mathfrak{k} \left([ H_B, \, \rho_{d,\mathfrak{k}} ] \right) \cap  {\cal F}_\mathfrak{k} \left( \rho  \right) \neq 0$ then just apply \eqref{eq:supp-ad-comm}. 
If, instead, we are in the case $  {\cal F}_\mathfrak{h} \left([ H_B, \, \rho_{d,\mathfrak{k}} ] \right) \cap  {\cal F}_\mathfrak{h} \left( \rho  \right) \neq 0$, then the only useful term in the expansion \eqref{eq:WequalKalman} is the last one, but this is enough to prove the claim.
The genericity of the argument can be shown as above.
\qed

\proof (of Theorem~\ref{thm:conver-trajtr})  
Consider the set $ {\cal N}$.
We want to show that the largest invariant set $ {\cal E} $ in $ {\cal N} $ is given by $ \rho_d $ only.
Condition~\ref{Ass1-thm1} guarantees that locally around $ \rho_d (t) $ there is no other equilibrium point in $ {\cal N} $, as, from Lemma~\ref{lem:Kalm-equiv}, the linearization at $ \rho_d $ is controllable.
Hence $ \rho_d $ is a locally asymptotically stable equilibrium for the closed loop system and $ \rho_d $ is isolated in ${\cal N}$.
Consider $ \rho_e \in {\cal N} $, $ \rho_e \neq \rho_d $. 
This implies $ \rho_e $ disjoint from $ \rho_d $ and $ V(\bmrho_d, \bmrho_e ) > 0$.
We need to show that $ \rho_e $ must be a repulsive equilibrium for the closed loop system \footnote{Since $ \bmrho_e $ may not be isolated in $ {\cal N}$, the term repulsive has to be intended as ``semi-repulsive''.}.
For $ \bmrho_e $ which is an antipodal point of a pure state $ \bmrho_d $, this is follows from Corollary~\ref{cor:equil-antipod-max}, since $ V (\bmrho_d, \bmrho_e ) $ is maximal in $ {\cal S}  $ while $ \dot{V} \leqslant 0$.
For any other $ \rho_e \in {\cal N} $, it is enough to perturb $ \rho_e $ to $ \tilde{\rho}_e \in {\cal S} $ so that $ {\cal F} \left([ H_B, \, \rho_d ] \right) \cap  {\cal F} \left( \tilde{\rho}_e \right) \neq 0$.
It is always possible to do this in a neighborhood of $\rho_e $ since $ {\cal F} \left([ H_B, \, \rho_d ] \right) $ has cardinality at least $ m/2 $ and $ ( h_{B, \Re, j\, j+1},  h_{B, \Im, j\, j+1}) \neq ( 0, \, 0)$ implies that $ {\rm Graph} (H_B) $ is connected and that there is no subspace $ \mathfrak{k}_{j\ell}$ invariant under ${\bm B}$.
But then, from Lemma~\ref{lem:support-equiv}, $ \llangle \bmrho_d,{\bm B}\tilde{\bmrho}_e \rrangle\neq 0 $ and $ V ( \bmrho_d, \tilde{\bmrho}_e ) <  V ( \bmrho_d, \bmrho_e ) $, i.e., $ \tilde{\rho}_e $ is attracted to $ \rho_d $.
To show that $ V ( \bmrho_e, \tilde{\bmrho}_e )$ increases, assume by contradiction that 
\beq
\dot{V} ( \bmrho_e, \tilde{\bmrho}_e ) = \llangle \dot \bmrho_e, \tilde{\bmrho}_e \rrangle + \llangle \bmrho_e, \dot{\tilde{\bmrho}}_e \rrangle = - \llangle \bmrho_d,{\bm B}\tilde{\bmrho}_e \rrangle  \llangle \bmrho_e,{\bm B}\tilde{\bmrho}_e \rrangle < 0. 
\label{eq:dotV-proof-Thm1}
\eeq
Consider the geodesic line in $ {\cal S}  $ connecting $ \bmrho_d $ with $ \bmrho_e $: $ \bmrho_\phi (s) = \bmrho_d + {\bm \phi} (s) $ such that $ {\bm \phi} (0) = 0 $ and $ {\bm \phi} (s_e ) = \bmrho_e - \bmrho_d $.
Along this line,
\[
\dot{V} ( \bmrho_\phi (s) , \tilde{\bmrho}_e ) = - \llangle \bmrho_d,{\bm B}\tilde{\bmrho}_e \rrangle  ^2 -  \llangle \bmrho_d,{\bm B}\tilde{\bmrho}_e \rrangle  \llangle {\bm \phi} (s) ,{\bm B}\tilde{\bmrho}_e \rrangle, \qquad s \in [0, s_e]
\]
is a function linear in $ {\bm \phi} (s) $ and such that, by the assumption \eqref{eq:dotV-proof-Thm1},
\beqan
\dot{V} ( \bmrho_\phi (0) , \tilde{\bmrho}_e ) & = & \dot{V} ( \bmrho_d , \tilde{\bmrho}_e ) = -  \llangle \bmrho_d,{\bm B}\tilde{\bmrho}_e \rrangle  ^2 < 0 \\
\dot{V} ( \bmrho_\phi (s_e) , \tilde{\bmrho}_e ) & = & \dot{V} ( \bmrho_e , \tilde{\bmrho}_e ) < 0.
\eeqan
But then $ \dot{V} ( \bmrho_\phi (s) , \tilde{\bmrho}_e ) < 0 $ $ \forall \; s\in [0, s_e] $, $ \dot{V} ( \bmrho_\phi (s) , \bmrho_\phi (s) ) =0$, meaning that $ \tilde{\bmrho}_e $ is attracted to the entire geodesic segment $ \bmrho_\phi(s) $, $  s \in [0, s_e] $, which is a contradiction, since $ \bmrho_d $ is an isolated equilibrium point.
Hence it must be $ \dot{V} ( \bmrho_e, \tilde{\bmrho}_e )\geqslant 0 $ i.e., $\bmrho_e $ is a repulsive equilibrium point.
Therefore $ \bmrho_e $ cannot belong to $ {\cal E}$.
From Lemma~\ref{lem:support-equiv}, all conditions \eqref{eq:deru-k-times} are satisfied or violated simultaneously respectively when $ \dot V =0 $ or $ \dot V <0 $, i.e., when $ {\cal F} \left([ H_B, \, \rho_d ] \right) \cap  {\cal F} \left( \tilde{\rho}_e \right) = 0$ or $ \neq 0$.
Hence outside $ {\cal N} $ the Jurdjevic-Quinn condition applies and $ \rho(0)$ must converge to $ \rho_d(t) $ since any other $ \rho_e \in {\cal N} $ is repulsive.
\qed

\begin{remark}
Condition~\ref{Ass1-thm1} of Theorem~\ref{thm:conver-trajtr} is obviously a necessary condition for convergence.
Condition~\ref{Ass2-thm1} instead is sufficient but not necessary, see Example 1 in Section~\ref{sec:cases}.
\end{remark}

While, from Lemma~\ref{lem:Kalm-equiv}, the linear span at $ \bmrho_d $ of the linearized system and of the $ \mathfrak{W}^\alpha $ yield a space of the same dimension, Item~\ref{en:item3-supp-equiv} of Lemma~\ref{lem:support-equiv} holds for the bilinear forms but it is in general not true for the linearization.

\begin{corollary}
\label{cor:rank-lin-bilin}
For $ H_A $ strongly regular:
\begin{enumerate}
\item $ {\rm rank }\; \mathfrak{W}^{\alpha-1} \bmrho_d ={\rm rank } \left[ {\bm  b} \,  {\bm A}{\bm  b } \ldots  {\bm A}^{\alpha-1} {\bm  b} \right]  , \qquad \forall \; \alpha = 0, \ldots, m-1 $;
\item $ \llangle \bmrho_d , [ \underbrace{{\bm A}, \ldots [ {\bm A}}_{\text{ $\alpha$ times }},{\bm B}]\ldots ]  \bmrho \rrangle  \neq 0 \; \; \not{\! \! \! \! \Longleftrightarrow} \llangle \bmrho ,{\bm A}^\alpha {\bm b} \rrangle  \neq 0 $
\end{enumerate}
\end{corollary}
\proof
The first point follows from the strong regularity of $ H_A $ and from \eqref{eq:supp-ad-comm}, which implies that the maximum number of independent vectors in the two sequences above is the same for all $ \alpha $.
The second from \eqref{eq:WequalKalman} and $  {\bm A} ^\alpha \bmrho_\mathfrak{h} =0 $.
\qed

The consequence is that the linearization alone is inconclusive about the region of attraction of the reference trajectory in the closed loop system, while instead the $ \ad$-commutators completely specify it.

\begin{corollary} 
\label{cor:dom-attr}
When $ {\rm Card} {\cal F}_\mathfrak{k} \left([ H_B, \, \rho_d ] \right) \geqslant m/2 $, the region of attraction of the system \eqref{eq:Liouv-vect-coh1} with the feedback \eqref{eq:feed-N-lev1} is given by $ {\cal R} = {\cal S} \setminus {\cal N}$.
\end{corollary}

\subsection{Global stabilization and topological obstructions} 
\label{sec:glob-top-ob}
The notions from differential topology used in this Section are recalled in Appendix~\ref{app:top-Morse}.
A compact manifold like $ {\cal S}  $ cannot be globally asymptotically stabilized because it lacks the contractivity property, i.e., it is not homotopy equivalent to a point, see \cite{Bhat1}, Proposition 1 and Theorem 1, and \cite{Wilson1}.
Proposition~\ref{prop:equil-antipod} suggests that for $ {\cal S}  $ this is due to the antipodal points.

\begin{proposition}
\label{prop:top-obstr-antip}
For $ {\cal S}  $, the $ \chi ({\cal S}  ) -1 $ antipodal points are irremovable topological obstruction to global stabilizability by smooth feedback.
\end{proposition}

\proof
Contractivity is a necessary condition for global asymptotic stabilizability.
For example, that $ {\cal S} = \mathbb{S}^2 $ with a point removed is homeomorphic (and hence homotopy equivalent) to $ \mathbb{R}^2 $ is well-known through the stereographic projection (see e.g. \cite{Armstrong1}, p. 34).
Since it is known that the domain of attraction of an asymptotically stable point must be homotopy equivalent to $ \mathbb{R}^m $ for some $m$ \cite{Wilson1}, then for $N=2 $ this is enough to affirm that convergence can be rendered global up to the antipodal point.
For $N>2 $, in order to show that the antipodal points are all obstructions to contractivity of $ {\cal S} $, consider the equilibrium $ \rho_d $ and one of its antipodal points $ \rho_{p_1}$.
By suitable change of basis, $ \rho_d $ and $ \rho_{p_1} $ can be rendered diagonal simultaneously.
By the transversality of the coadjoint orbit on $ \mathfrak{h}$, it is possible to determine a submanifold of $ {\cal S}  $ connecting $ \rho_d $ and $ \rho_{p_1} $ and not passing through any other of the antipodal points. 
To see it, notice that for $N>2 $ it is always possible to adjust the basis \eqref{eq:basis-elem1}-\eqref{eq:basis-elem4} so as to attain a 3-dimensional simple subalgebra of $ \mathfrak{su}(N)$, which, as described in Section~\ref{sec:topol-descr}, draws an $ \mathbb{S}^2 $ orbit under the adjoint action (see \cite{Sanchez1} for the details of this construction).  
This is a well-defined compact submanifold of $ {\cal S}  $ and it is not contractible for what said above for $N=2$. 

\qed

\section{A few cases of physical interest}
\label{sec:cases}
The methods developed above yield considerable insight into the stabilizability and convergence properties of a quantum density operator. A few interesting cases for $ N$-level systems are now described. 
It is followed by a more detailed description for systems with $N=2, \, 3 $.
\begin{itemize}
\item Since $ m\leqslant N^2 - N $, and $ {\rm Card} {\cal F}_\mathfrak{k} \left( [ H_B, \rho_d ] \right) \leqslant (N^2 - N)/2$, (i.e., the maximal number of off-diagonal terms), each complex flag manifold $ {\cal S}  $ may admit a controllable linearization (depending on $ \rho_d $).
\item The assumption of direct coupling between nearest energy levels $  ( h_{B, \Re, j\, j+1},  h_{B, \Im, j\, j+1}) \neq ( 0, \, 0)$, is needed in order to exclude the existence of subset of $ {\cal S} $ which remains invariant under the closed loop dynamics.
It is a common assumption in most practical cases (dipole approximation \cite{Dahleh1}). 
See also Example 2 below (last item).
\item The full connectivity of $ {\rm Graph} (H_B) $ is neither a sufficient nor a necessary condition for asymptotic stability.
\item
If $ \rho_d $ is an eigenstate and $ \rho $ another eigenstate then there is never convergence, not even if $ {\rm Graph}(H_B ) $ is fully connected, because $ \rho $ is antipodal to $ \rho_d $.

\item 
For pure states and not fully connected $ {\rm Graph} (H_B)$, certain eigenstates are easier to stabilize than others. 
The easiest is the one of energy $ {\cal E}_j $ such that the index $ j$ appears more often in $ {\cal F} _\mathfrak{k} \left( [ H_B, \rho_d ]  \right) $. In Example 2 below with $ H_B $ in \eqref{eq:3lev-H_B1}, it is easier to stabilize to the eigenstate of intermediate energy than to the ground state or to the most excited state.
When $ {\rm Graph} (H_B)$ is fully connected, there is no such difference.
From Theorem~\ref{thm:conver-trajtr}, this does not mean that all initial conditions have the same convergence properties to a given $ \rho_d $.

\item
If $ \rho_d $ and $ \rho(0) $ are both block diagonal and the blocks do not overlap
\[
\rho_d =\begin{bmatrix} 
\begin{matrix} * & \hdots & *  \\ \vdots & & \vdots  \\ * & \ldots & * \\ \hline \end{matrix} \; \vline  & \\
&\! \!  \begin{matrix} &  & &  & &   \\  &  & &   \\ &  & &   \end{matrix} 
\end{bmatrix}, \quad 
\rho(0) =\begin{bmatrix} 
\begin{matrix} &  & &  & &   \\  &  & &   \\ &  & &  \end{matrix} 
& \\
&\! \!\! \! \vline\;  \begin{matrix}\hline * & \hdots & *  \\ \vdots & & \vdots  \\  * & \ldots & *  \end{matrix} 
\end{bmatrix},
\]
then
\[
[H_B, \, \rho_d ] =\begin{bmatrix} 
\begin{matrix} *  & \hdots  &   *   \\ \vdots  &   &   \\ * & \hdots & * \end{matrix} &
\begin{matrix} *  &   \hdots & *  \\   &  &  \vdots \\ * & \hdots & *  \end{matrix}  \\
\begin{matrix} * & \hdots & * \\ \vdots  &   & \vdots  \\ * & \hdots   & *  \end{matrix}
\; &
\! \!\! \! \vline\;  
\begin{matrix}\hline 0 & \hdots & 0  \\ \vdots & & \vdots  \\  0 & \ldots & 0  \end{matrix} 
\end{bmatrix},
\]
which implies $ {\cal F} \left([ H_B, \, \rho_d ] \right) \cap  {\cal F} \left( \rho (0) \right) = 0$ and $ \dot V =0 $, i.e., $ \rho(0)$ is not attracted to $ \rho_d$.
Since $ \tr{ \rho \rho_d} =0 $, $ \rho_d $ and $ \rho $ are as distant as antipodal states.

\item
Not all states in  ${\cal N} $ are maximally distant from $ \rho_d $.
Assume $ \rho_d , \rho $ such that $ {\cal F}_\mathfrak{h} \left([ H_B, \, \rho_d ] \right) \cap  {\cal F}_\mathfrak{h}  \left( \rho (0) \right) \neq 0$, $  {\cal F}_\mathfrak{k}  \left( \rho_d \right) =0$, $ {\cal F}_\mathfrak{k} \left( H_B \right) \cap  {\cal F}_\mathfrak{k}  \left( \rho  \right) = 0$.
Also in this case $ {\cal F} \left([ H_B, \, \rho_d ] \right) \cap  {\cal F} \left( \rho (0) \right) = 0$ and $ \rho $ is not converging.
However, since $ \tr{ \rho_d \rho} \neq 0 $, $ \rho_d $ and $ \rho$ are not maximally distant.
\item A typical example of initial condition such that $ \dot{V} ( \bmrho_d (0) , \bmrho (0) ) =0 $ but not invariant under the drift (see paragraph before Lemma~\ref{lem:support-equiv}) is attained when  $ {\cal F}_\mathfrak{k} \left([ H_B, \, \rho_d ] \right) \cap  {\cal F}_\mathfrak{k}  \left( \rho (0) \right) \neq 0$ but $ [ H_B, \, \rho_d (0) ] $, $ \rho(0)$ both real or purely imaginary.
This follows from Proposition~\ref{cor:diag-comm1}.

\end{itemize}

\noindent
{\bf Example 1 (cont'd) } 
Assume 
\[
H_A = \frac{h_{A,\mathfrak{h},1} }{\sqrt{2}} \begin{bmatrix} 1 & 0 \\ 0 & -1 \end{bmatrix} = h_{A,\mathfrak{h},1}  \lambda_{\mathfrak{h},1} \quad
\text{ and }  \quad
H_B = \frac{h_{B,\mathfrak{k},\Re,12} }{\sqrt{2}}  \begin{bmatrix} 0 & 1 \\ 1 & 0 \end{bmatrix} = h_{B,\mathfrak{k},\Re,12} \lambda_{\mathfrak{k},\Re,12} 
\]
Then 
\[
{\bm A} = -i \sqrt{2} h_{A,\mathfrak{h},1}  \ad_{\lambda_{\mathfrak{h},1}} = 2 h_{A,\mathfrak{h},1}\begin{bmatrix} 0 & -1 & 0 \\ 1 & 0 & 0 \\ 0 & 0 & 0 \end{bmatrix}  \quad
\text{ and } \quad
{\bm B} = -i \sqrt{2} h_{B,\mathfrak{k},\Re,12}   \ad_{\lambda_{\mathfrak{k},\Re,12}} = 2 h_{B,\mathfrak{k},\Re,12} \begin{bmatrix} 0 & 0 & 0 \\ 0  & 0 & -1  \\ 0& 1 & 0  \end{bmatrix} 
\]
From Proposition~\ref{cor:diag-comm1}, both $ \| \bmrho_\mathfrak{k} \| $ and $ \bmrho_\mathfrak{h} $ are integrals of motion of the unforced dynamics, while the two components of $ \bmrho_\mathfrak{k}$ evolve according to a sinusoidal law.
When applying Theorem~\ref{thm:conver-trajtr} to the system plus the feedback \eqref{eq:feed-N-lev1}, we have the following for the closed loop system:
\begin{itemize}
\item any $ \rho_d $ has a single antipodal point which also is an equilibrium;
\item if $ \rho_d $ diagonal, $ {\cal F}_\mathfrak{k} \left( [ H_B, \rho_d ] \right) = \{ (12) \} $, the linearization is controllable and any nondiagonal $ \rho $ satisfies Theorem~\ref{thm:conver-trajtr}.
Hence any $ \rho(0) $ such that $ \rho_\mathfrak{k} (0) \neq 0 $ is attracted to $ \rho_d $ diagonal;
\item if $ \rho_d $ off-diagonal, $ {\rm Card} {\cal F}_\mathfrak{k} \left( [ H_B, \rho_d ] \right) = 0 $, and the sufficient condition of Theorem~\ref{thm:conver-trajtr} does not apply.
However, $ {\cal F}_\mathfrak{h} \left( [ H_B, \rho_d ] \right) \neq 0 $ and as long as $  {\cal F}_\mathfrak{h} \left( [ H_B, \rho_d ] \right) \cap {\cal F}_\mathfrak{h} \left( \rho(0) \right) \neq 0 $, i.e., whenever $ \rho_\mathfrak{h} \neq 0 $, $ \rho \to \rho_d$.
This is a special situation due to $ {\rm dim} ( \mathfrak{h}) =1 $, and has no counterpart for $ N> 2$.
\end{itemize}
In summary, there is always almost global convergence except when $ \bmrho_{d,\mathfrak{h}} = \bmrho_\mathfrak{h} =0 $, i.e., except when both $ \bmrho_d $ and $ \bmrho $ belong to great horizontal circles.
\qed

\noindent
{\bf Example 2 (cont'd) } 
The drift of the system is given by
\[
H_A = \frac{h_{A,\mathfrak{h},1} }{\sqrt{2}}  \begin{bmatrix} 1 & 0 & 0 \\ 0 & -1 & 0 \\ 0 & 0 & 0  \end{bmatrix} + \frac{h_{A,\mathfrak{h},2} }{\sqrt{6}}  \begin{bmatrix} 1 & 0 & 0 \\ 0 & 1 & 0 \\ 0 & 0 & -2  \end{bmatrix}  = h_{A,\mathfrak{h},1} \lambda_{\mathfrak{h},1} +  h_{A,\mathfrak{h},2} \lambda_{\mathfrak{h},2} .
\]
We shall consider the following control vector field
\beq
H_B =  \frac{1}{\sqrt{2}} \begin{bmatrix} 0 &  h_{B,\mathfrak{k},\Re,12}  & 0 \\  h_{B,\mathfrak{k},\Re,12}  & 0 &  h_{B,\mathfrak{k},\Re,23}  \\ 0 &  h_{B,\mathfrak{k},\Re,23}  & 0  \end{bmatrix} =  h_{B,\mathfrak{k},\Re,12}  \lambda_{\mathfrak{k},\Re,12} + h_{B,\mathfrak{k},\Re,23}  \lambda_{\mathfrak{k},\Re,23} ,
\label{eq:3lev-H_B1}
\eeq
which has $ {\cal F}_\mathfrak{k} ( H_B ) = \{ (12), (23) \} $ or, alternatively, 
\beq
H_B =  \frac{1}{\sqrt{2}} \begin{bmatrix} 0 &  h_{B,\mathfrak{k},\Re,12}  & h_{B,\mathfrak{k},\Re,13} \\  h_{B,\mathfrak{k},\Re,12}  & 0 &  h_{B,\mathfrak{k},\Re,23}  \\ h_{B,\mathfrak{k},\Re,13} &  h_{B,\mathfrak{k},\Re,23}  & 0  \end{bmatrix} =  h_{B,\mathfrak{k},\Re,12}  \lambda_{\mathfrak{k},\Re,12} + h_{B,\mathfrak{k},\Re,13}  \lambda_{\mathfrak{k},\Re,13} + h_{B,\mathfrak{k},\Re,23}  \lambda_{\mathfrak{k},\Re,23}
\label{eq:Hc-full-3lev}
\eeq
which has a ``fully connected'' graph, $   {\cal F}_\mathfrak{k} (H_B ) = \{ (12), (13),(23) \} $.

A list of interesting cases is the following:

\begin{itemize}
\item any of the (two for pure/psudopure, five for the generic case) antipodal points of any $ \rho_d $ is also an equilibrium.
\item $ \rho_d $ diagonal: only the off-diagonal part of $ \rho $ matters
\begin{itemize}
\item $ \rho_d $ pure (or pseudopure), e.g. $ \rho_d = \rho_{v_1}$
\begin{itemize} 
\item $ H_B $ given in \eqref{eq:3lev-H_B1}: $ {\cal F} _\mathfrak{k} \left( [ H_B, \rho_d ]  \right) = \{ (12) \}$ $ \Longrightarrow $ the linearization is never controllable since $ 2 \, {\rm Card}{\cal F} _\mathfrak{k} \left( [ H_B, \rho_d ]  \right)< 4 =m $, hence Theorem~\ref{thm:conver-trajtr} does not apply.
Unlike the $N=2 $ case, now in general $ \rho(0) \not{\! \! \to} \rho_d$; 
\item $ H_B $ given in \eqref{eq:Hc-full-3lev}: $ {\cal F} _\mathfrak{k} \left( [ H_B, \rho_d ]  \right) = \{ (12), (13) \}$ $ \Longrightarrow $ the linearization is controllable. Any $ \rho(0) $ such that $ {\cal F} _\mathfrak{k} \left(\rho(0) \right) \cap \{ (12), (13) \} \neq 0 $ is converging.
However, if one considers the pure state
\[ 
\rho(0) =  \frac{1}{2} \begin{bmatrix} 0 & 0 & 0 \\ 0 & 1 & 1 \\ 0 & 1 & 1 \end{bmatrix} ,
\] 
then $ {\cal F} _\mathfrak{k} \left( \rho \right) = \{ (23) \}$, implying $ \dot V (0) = u = \llangle \bmrho_d (0) , \,{\bm B}\bmrho(0) \rrangle =0 $, i.e., the system is not converging to $ \rho_d $ in spite of the Kalman controllability condition on the linearization.
Notice how for this example $ {\rm rank} \left( \mathfrak{W}^3 \bmrho_d \right) ={\rm rank} \left( \mathfrak{W}^3 \bmrho(0) \right) = 4$, while $ \bmrho_d^T  \mathfrak{W}^3 \bmrho(0) = \{ 0,0,0,0\}$.

\end{itemize}
\item $ \rho_d $ pure (or pseudopure), but $ \rho_d = \rho_{v_2}$
\begin{itemize} 
\item $ H_B $ either \eqref{eq:3lev-H_B1} or \eqref{eq:Hc-full-3lev}: $ {\cal F} _\mathfrak{k} \left( [ H_B, \rho_d ]  \right) = \{ (12), (23) \}$ $ \Longrightarrow $ the linearization is always controllable. Any $ \rho(0) $ such that $ {\cal F} _\mathfrak{k} \left(\rho(0) \right) \cap \{ (12), (23) \} \neq 0 $ is converging.
\end{itemize}
\item $ \rho_d $ with all different eigenvalues, e.g. $ \rho_d = \rho_{g_1}$
\begin{itemize} 
\item $ H_B $ in \eqref{eq:3lev-H_B1}: $ {\cal F} _\mathfrak{k} \left( [ H_B, \rho_d ]  \right) = \{ (12), (23) \}$ $ \Longrightarrow $ the linearization is never controllable since now $ m=6 $; 
\item $ H_B $ in \eqref{eq:Hc-full-3lev}: $ {\cal F} _\mathfrak{k} \left( [ H_B, \rho_d ]  \right) = \{ (12), (13), (23) \}$ $ \Longrightarrow $ the linearization is always controllable. Any $ \rho(0) $ such that $ {\cal F} _\mathfrak{k} \left(\rho(0) \right) \neq 0 $ is converging, any $  \rho(0) $ such that $ {\cal F} _\mathfrak{k} \left(\rho(0) \right) = 0 $ is antipodal.
\end{itemize}
\end{itemize}
\item $ \rho_d - \varrho_0 \lambda_0 $ off-diagonal
\begin{itemize} 
\item $ H_B $ in \eqref{eq:3lev-H_B1} and $  {\cal F} _\mathfrak{k} \left( \rho_d   \right) \subseteq {\cal F} _\mathfrak{k} \left( H_B   \right) $ $ \Longrightarrow $ linearization is never controllable, hence Theorem~\ref{thm:conver-trajtr} does not apply and in general $ \rho(0)  \not{\! \! \to} \rho_d(t)$;
\item $ H_B $ in \eqref{eq:3lev-H_B1} and $  {\cal F} _\mathfrak{k} \left( \rho_d   \right) \not{\! \! \subseteq} {\cal F} _\mathfrak{k} \left( H_B   \right) $ $ \Longrightarrow $ ${\rm Card} {\cal F} _\mathfrak{k} \left( [ H_B, \rho_d ]  \right) $ is at least 2, implying that the linearization is controllable at least for pure/pseudopure states;
\item if $  {\cal F} _\mathfrak{k} \left( \rho_d   \right) \cap {\cal F} _\mathfrak{k} \left( H_B   \right) \neq 0$, then also $ {\cal F} _\mathfrak{h} \left( \rho (0)   \right) $ matters for the convergence, see \eqref{eq:comm-h1};
\item if $  {\cal F} _\mathfrak{k} \left( \rho_d   \right) \cap {\cal F} _\mathfrak{k} \left( H_B   \right) = 0$ then convergence depends only on $ {\cal F} _\mathfrak{k} \left( \rho (0)   \right) $ (plus controllability), see \eqref{eq:comm-kk}.
\end{itemize}
\item If the control Hamiltonian is $
H_B =   h_{\mathfrak{k},\Re,12}  \lambda_{\mathfrak{k},\Re,12} + h_{\mathfrak{k},\Re,13}  \lambda_{\mathfrak{k},\Re,13} $, i.e., direct coupling between $ {\cal E}_2 $ and $ {\cal E}_3 $ is missing, then the sufficient condition of Theorem~\ref{thm:conver-trajtr} does not apply.
Assume for example 
\[ 
\rho_d =   \begin{bmatrix} 0 & 0 & 0 \\ 0 & \ast & \ast \\ 0 & \ast & \ast \end{bmatrix} , \qquad
\rho(0) =   \begin{bmatrix} \ast & \ast & 0 \\ \ast & \ast & 0 \\ 0 & 0 & 0 \end{bmatrix} .
\] 
Then $ {\cal F} _\mathfrak{k} \left( [ H_B, \rho_d ]  \right) = \{ (12), (13) \} $ and $ {\cal F} _\mathfrak{k} \left( [ H_B, \rho_d ]  \right) \cap {\cal F} _\mathfrak{k} \left( \rho (0)  \right) = \{ (12) \} $. 
However, \mbox{$ \rho(0) \not{\! \! \to} \rho_d(t)$}.
\end{itemize}
\qed

\section{Acknowledgments}
The author would like to thank A. Agrachev and P. Rouchon for discussion on the topic of this work.

\appendix

\section{On the adjoint representation}
\label{sec:adj-rep}

A {\em representation} of a Lie algebra $ \mathfrak{g} $ on a vector space $ {\cal X} $ is a mapping $ \Theta \, : \; \mathfrak{g} \to \mathfrak{gl} ({\cal X}) $ which is a Lie algebra homomorphism, i.e., a map which 
\begin{enumerate}
\item is linear $ \Theta (\alpha_1 A_1 + \alpha_2 A_2 ) = \alpha_1 \Theta ( A_1  ) + \alpha_2 \Theta (  A_2 ) $, $ \forall \; A_1 , \, A_2 \in \mathfrak{g}$ and $ \forall \; \alpha_1 , \, \alpha_2 $ in the field of $ {\cal X} $;
\item preserves the Lie bracket $  \Theta ([ A_1 ,\, A_2 ] ) =  [ \Theta (A_1  ), \,  \Theta (A_2  )] $, $ \forall \; A_1 , \, A_2 \in \mathfrak{g}$.
\end{enumerate}

So a representation $ \Theta $ assigns to each $ A\in \mathfrak{g} $ a linear operator  $\Theta (A) \in \mathfrak{gl}({\cal X}) $.
A particularly useful representation is the adjoint representation on ${\cal X} = \mathbb{R}^n $.
If for an $n$-dimensional Lie algebra $ \mathfrak{g} $ we choose the basis $ A_1, \ldots A_n $ then the Lie brackets of the basis elements are $
 [ A_j, \, A_k ] = \sum_{\ell=1}^n c_{jk}^\ell A_\ell $.
The components of the 3-tensor $ c_{jk}^\ell $ are called structure constants of the Lie algebra with respect to the basis $ A_1, \ldots A_n $.
The {\em adjoint representation} of $ \mathfrak{g} $, $ {\rm ad}_\mathfrak{g} $, with respect to the basis $ A_1, \ldots A_n $ is the representation having as basis elements the $ n \times n $ matrices of structure constants $ {\bm A}_1, \ldots , {\bm A}_n $, $ {\bm A}_j = {\rm ad}_{A_j} = [ A_j , \, \cdot \, ]  $ of entries $ \left( {\bm A}_j \right)_{\ell k} =  c_{jk}^\ell $.
Notice how the two free indexes $ k $ and $ \ell $ identify respectively the columns and the rows of the new basis elements.

In general, the adjoint representation of a linear Lie algebra is a derivation of the algebra and corresponds to the infinitesimal representation of all the one-parameter groups of automorphisms.
For a semisimple compact Lie algebra $ \mathfrak{g} $, the main features of $ \ad_\mathfrak{g} $ are (see e.g. \cite{Sattinger1}, p. 39 and 129):
\begin{itemize}
\item it is a real semisimple Lie algebra;
\item it is isomorphic to $ \mathfrak{g} $;
\item $ \forall \, A, \, B \in \mathfrak{g} $: $ [ \ad_A, \, \ad_B ] = \ad_{[ A, \, B ]}$.
\end{itemize}
Let us spend some more words on emphasizing how the ``linearity'' of the adjoint representation may be intended, which is one of the {\em leitmotifs} of the paper.
If $ B \in  \mathfrak{g} $ has the expression $ B = b_1 A_1 + \ldots b_n A_n $, then as long as we keep the basis fixed, $ B $ is uniquely identified by its vector of components: $
 B \simeq {\bm b} = \begin{bmatrix} b_1 & \ldots & b_n \end{bmatrix}^T$.
Then
\beq
[ A_j , \, B ] = [ A_j ,\,\sum_{k=1}^n b_k A_k ] \simeq  \sum_{k , \, \ell =1}^n {\rm e}_\ell^T \left( {\rm ad}_{A_j}\right)_{\ell k} b_k = {\bm A}_j {\bm b} .
\label{eq:lie-brack-adj1}
\eeq
We will often make the double substitution $ \{ A,\, B \}  \simeq \{  {\bm A},\, {\bm b}\}  $ which will correspond to replacing the (bilinear) matrix commutator $ [ \, \cdot\, , \, \cdot \, ] \; :  \mathfrak{g} \times \mathfrak{g}  \to \mathfrak{g} $ with the linear operation $ {\rm ad}_\mathfrak{g} \times \mathbb{R}^n   \to \mathbb{R}^n $, i.e., left matrix multiplication.

\section{A few facts from topology}
\label{app:top-Morse}
The material in this Appendix is taken from standard texts on (differential) topology e.g. \cite{Armstrong1,Guillemin1}.
Let $ {\cal X} $, $ {\cal Y} $ be topological spaces and $ f, \, g \,: {\cal X} \to {\cal Y} $ be continuous maps.
The mapping $ f $ is said {\em homotopic} to $ g$ if ``it can be continuously deformed to $ g$'', i.e., if $ \exists $ a continuous mapping $ h \,: {\cal X} \times [0, \, 1 ] \to {\cal Y}$ such that $ h( x, 0) = f(x) $, $ h(x, 1) = g(x) $ $ \forall \; x \in {\cal X} $.
$ {\cal X} $, $ {\cal Y} $ are said {\em homotopy equivalent} if $ \exists $ maps $ f \,: {\cal X} \to {\cal Y} $ and $  g \,: {\cal Y} \to {\cal X} $ such that $ g \circ f $ and $ f \circ g $ are homotopic to the identity maps in $ {\cal X} $ and $ {\cal Y }$ respectively.
$ {\cal Y} $ is said {\em contractible} if the identity map on $ {\cal Y} $ is homotopic to the constant map $ {\cal Y} \to x_\circ $ for any $ x_\circ\in {\cal Y} $.
A space is contractible if and only if it is homotopy equivalent to a point. 
No compact manifold is contractible. 
Homotopy equivalence is an equivalence relation on topological spaces and the classes of homotopy equivalent spaces are called homotopy types.
Examples used in this paper are:
\begin{itemize}
\item $ \mathbb{S}^m - \{ p \} $ is homotopy equivalent to $ \mathbb{R}^m $ ($ p $ is any point of $ \mathbb{S}^m $);  
\item $ \mathbb{R}^m - \{ 0 \} $ is homotopy equivalent to $ \mathbb{S}^{m-1} $ 
\item $ \mathbb{R}^m $ is homotopy equivalent to a point (and hence to any contractible space).
\end{itemize}

\bibliographystyle{abbrv} 
\small

\end{document}